\documentclass[aps, pre, twocolumn, a4paper, floatfix, nofootinbib, longbibliography]{revtex4-1}

\usepackage[utf8]{inputenc}
\usepackage{epsfig}
\usepackage[T1]{fontenc}
\usepackage[english]{babel}
\usepackage[table]{xcolor}
\usepackage{t1enc}
\usepackage{graphicx}
\usepackage{amssymb}
\usepackage{amsmath}
\usepackage{relsize}
\usepackage{overpic}
\usepackage{bm}

\usepackage[normalem]{ulem}

\newcommand{\avg}[1]{{\left<#1\right>}}

\usepackage{mathtools}
\def\multiset#1#2{\ensuremath{\left(\kern-.3em\left(\genfrac{}{}{0pt}{}{#1}{#2}\right)\kern-.3em\right)}}

\usepackage{amsmath}

\usepackage{multirow}

\usepackage{verbatim}
\usepackage{overpic}
\usepackage{booktabs}

\begin{document}

\title{Inferring the mesoscale structure of layered, edge-valued and time-varying networks}

\author{Tiago P. Peixoto}
\email{tiago@itp.uni-bremen.de}
\affiliation{Institut f\"{u}r Theoretische Physik, Universit\"{a}t Bremen, Hochschulring 18, D-28359 Bremen, Germany}


\begin{abstract}

Many network systems are composed of interdependent but distinct types
of interactions, which cannot be fully understood in isolation. These
different types of interactions are often represented as layers,
attributes on the edges or as a time-dependence of the network
structure. Although they are crucial for a more comprehensive scientific
understanding, these representations offer substantial
challenges. Namely, it is an open problem how to precisely characterize
the large or mesoscale structure of network systems in relation to these
additional aspects. Furthermore, the direct incorporation of these
features invariably increases the effective dimension of the network
description, and hence aggravates the problem of overfitting, i.e.  the
use of overly-complex characterizations that mistake purely random
fluctuations for actual structure. In this work, we propose a robust and
principled method to tackle these problems, by constructing generative
models of modular network structure, incorporating layered, attributed
and time-varying properties, as well as a nonparametric Bayesian
methodology to infer the parameters from data and select the most
appropriate model according to statistical evidence. We show that the
method is capable of revealing hidden structure in layered, edge-valued
and time-varying networks, and that the most appropriate level of
granularity with respect to the additional dimensions can be reliably
identified. We illustrate our approach on a variety of empirical
systems, including a social network of physicians, the voting
correlations of deputies in the Brazilian national congress, the global
airport network, and a proximity network of high-school students.

\end{abstract}

\maketitle

\section{Introduction}

The network abstraction has been successfully used as a powerful
framework behind the modeling of a great variety of biological,
technological and social
systems~\cite{newman_networks:_2010}. Traditionally, most network models
proposed in these contexts consist of a set of elements possessing a
single type of pairwise interaction (e.g. epidemic contact, transport
route, metabolic reaction, etc.). More recently, it has becoming
increasingly clear that single types of interaction do not occur in
isolation, and that a complete system encompasses several layers of
interactions~\cite{kivela_multilayer_2014, boccaletti_structure_2014,
de_domenico_mathematical_2013}, and very often change in
time~\cite{holme_temporal_2012}. Many examples have shown that the
interplay between different types of interactions can dramatically
change the outcome of paradigmatic processes such as
percolation~\cite{buldyrev_catastrophic_2010}, epidemic
spreading~\cite{dickison_epidemics_2012, chen_outbreaks_2013,
valdano_analytical_2015}, diffusion~\cite{gomez_diffusion_2013,
radicchi_abrupt_2013}, opinion formation~\cite{diakonova_absorbing_2014,
masuda_voter_2014, halu_connect_2013}, evolutionary
games~\cite{gomez-gardenes_evolution_2012, jiang_spreading_2013,
santos_biased_2014}, and
synchronization~\cite{boccaletti_structure_2014,
gambuzza_intra-layer_2014}, among others. The realization that different
types of interaction need to be incorporated into network models also
changes the way data need to be analyzed. In particular, the large or
mesoscale structure of network systems may be intertwined with the
layered or temporal structure, in such a way that cannot be visible if
this information is omitted. The conventional approach of representing
mesoscale structures is to separate the nodes into groups (or modules,
``communities'') that have a similar role in the network
topology~\cite{fortunato_community_2010}. Some methods have been
proposed to identify such groups in both
layered~\cite{mucha_community_2010,
de_domenico_mathematical_2013,bazzi_community_2014,
de_domenico_identifying_2015} and
time-varying~\cite{mucha_community_2010, rosvall_mapping_2010,
ronhovde_detecting_2011, bassett_robust_2013,bazzi_community_2014,
sarzynska_null_2014, gauvin_detecting_2014,macmahon_community_2015}
networks. However, these methods do not address two very central
questions: 1. Is the layered or temporal structure indeed important for
the description of the network?  And if so, to what degree of
granularity? 2. How does one distinguish between multiple descriptions
of the same network, and in particular separate actual structure from
stochastic fluctuations? In this work we tackle both these questions by
formulating generative models of layered networks, obtained by
generalizing several variants of the stochastic block
model~\cite{holland_stochastic_1983,fienberg_statistical_1985,
faust_blockmodels:_1992, anderson_building_1992}, incorporating features
such as hierarchical structure~\cite{clauset_hierarchical_2008,
peixoto_hierarchical_2014}, overlapping groups~\cite{airoldi_mixed_2008,
ball_efficient_2011, peixoto_model_2015} and
degree-correction~\cite{karrer_stochastic_2011}, in addition to
different types of layered structure. We show how the unsuspecting
incorporation of many layers that happen to be uncorrelated with the
mesoscale structure can in fact hinder the detection task, and obscure
structure that would be visible by ignoring the layer division in the
usual fashion. Since most methods proposed so far take any available
layer information for granted, and attempt to model it in absolute
detail, this issue represents a severe limitation of these methods in
capturing the structure of layered networks in a reliable manner.  We
show how this problem can be solved by performing model selection under
a general nonparametric Bayesian framework, that can also be used to
select between different model flavors (e.g. with overlapping groups or
degree correction). We demonstrate that the proposed methodology can
also be used to infer mesoscale structure in networks with real-valued
correlates on the edges (such as weights, distances, etc.), while
reliably distinguishing structure from noise, as well as change-points
in time varying networks~\cite{peel_detecting_2014}.

This work extends recent developments on
layered~\cite{valles-catala_multilayer_2014, jiang_stochastic_2015,
barbillon_stochastic_2015, paul_community_2015,corneli_exact_2015,
stanley_clustering_2015}, edge-valued~\cite{mariadassou_uncovering_2010,
guimera_network_2013, rovira-asenjo_predicting_2013,
aicher_learning_2014} and temporal~\cite{yang_detecting_2010,
xu_dynamic_2013, xu_dynamic_2014, xu_stochastic_2014,
ghasemian_detectability_2015} generative processes, not only by
incorporating many important topological patterns simultaneously
(i.e. hierarchical structure, degree correction and overlapping groups),
but also by tying all these types of model into a nonparametric Bayesian
framework that permits model selection, and avoids overfitting. The
framework presented allows one not only to select among all different
model classes, but also their appropriate order, i.e. the number of
groups, layer bins and hierarchical structure. This is done in a
principled fashion, based on statistical evidence and the principle of
parsimony, and without the specification of \emph{ad hoc}
parameters. Furthermore, since it is based on the computation of
posterior probabilities, it can be extended to other probabilistic
models.

This paper is divided as follows. In Sec.~\ref{sec:model} we formulate
generative models for layered structure, including a very diverse set of
possible topological patterns, and in Sec.~\ref{sec:selection} we
describe a Bayesian model selection procedure to choose between them
based on statistical evidence. In Sec.~\ref{sec:info} we tackle the
problem of deciding whether or not the layered structure is informative
of the network structure. In Sec.~\ref{sec:real} we show how the layered
models can be adapted to networks with real-valued edge-covariates, and
in Sec.~\ref{sec:time} to networks that change in time, for which the
division into layers corresponds to a detection of change-points. We
finalize in Sec.~\ref{sec:conclusion} with a conclusion.

\section{Generative models of layered networks}\label{sec:model}

We consider graphs that have a layered
structure~\cite{boccaletti_structure_2014,kivela_multilayer_2014}, so
that the adjacency matrix in layer $l \in [1, C]$ can be written as
$A_{ij}^l$ (with values in the range $[0, 1]$ for a simple graph, or in
$\mathbb{N}$ for a multigraph), corresponding to the presence of an edge
between vertices $i$ and $j$ in layer $l$. We will consider both
directed and undirected graphs (i.e. $A_{ij}^l$ being asymmetric and
symmetric, respectively), although we will focus on the undirected case
in most of the derivations, since the directed cases are mostly
straightforward modifications (which are summarized in
Appendix~\ref{app:directed}). Here we assume that the vertices are
globally indexed, and in principle can receive edges in all layers. The
\emph{collapsed graph} corresponds to the merging of all edges in a
single layer, with a resulting adjacency matrix
$A_{ij}=\sum_lA_{ij}^l$. In the following, we will denote a specific
layered graph as $\{G_l\}$ (with $G_l=\{A_{ij}^l\}$ being an individual
layer), and its corresponding collapsed graph as $G_c=\{A_{ij}\}$.

\begin{figure}
  \begin{tabular}{cc}
    \begin{minipage}{0.49\columnwidth}\centering
      \includegraphics{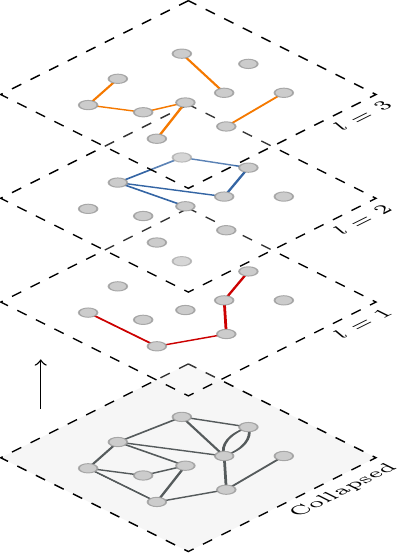}\\
      Edge covariates
    \end{minipage}  &
    \begin{minipage}{0.49\columnwidth}\centering
      \includegraphics{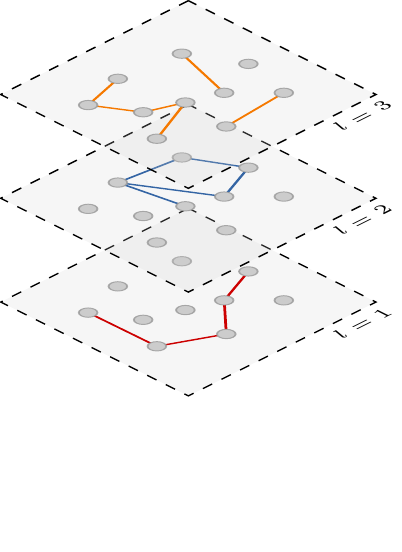}\\
      Independent layers
    \end{minipage}

  \end{tabular}

  \caption{(Color online) Two processes capable of generating layered
  networks. \emph{Left:} The collapsed graph is generated first, and
  conditioned on it, the edges are distributed among the
  layers. \emph{Right:} The layers are formed independently from each
  other.\label{fig:generation}}
\end{figure}

In this work we will consider two alternative ways of generating a given
layered graph $\{G_l\}$ (see Fig.~\ref{fig:generation}). The first
approach interprets the layers as edge
covariates~\cite{mariadassou_uncovering_2010}: First the collapsed graph
$G_c$ is generated, and then the layer membership of each edge is
 a random variable sampled from a distribution conditioned on the
adjacent vertices. In the second approach, the graphs $G_l$ at each
layer $l$ are generated independently from each other. (Henceforth we
call these alternatives simply by ``edge covariates'' and ``independent
layers'', respectively). These different generative processes do not
exhaust the realm of possible multilayer models. Instead, the objective
here is to consider the most basic possibilities that allow us to
incorporate different types of properties into the generated networks,
and enable the formulation of a nonparametric model selection framework
to decide if either one is more appropriate than the other depending on
the statistical evidence available in the data, as discussed in detail
below.

In the following we define two versions of the stochastic block model
family (SBM), corresponding to the alternatives outlined above.

\subsection{SBM with edge covariates}

We generate first a collapsed graph from the traditional SBM ensemble,
where $N$ nodes are divided into $B$ groups, via the membership vector
$\{b_i\}\in [1, B]^N$, and the number of edges randomly placed between
groups $r$ and $s$ is given by the edge counts $e_{rs}$ (or twice the
number if $r=s$, for convenience of notation). After the graph is
generated, for each set of edges incident on groups $r$ and $s$, we
distribute the layer memberships randomly, conditioned only on the total
number of edges of each type $l$ between the two groups, $m_{rs}^l$. Any
particular distribution of covariates among edges incident on groups $r$
and $s$ is generated with the same probability, which in the case of
simple undirected graphs is given by
\begin{equation}
  \frac{\prod_lm_{rs}^l!}{m_{rs}!},
\end{equation}
where $m_{rs} = \sum_lm_{rs}^l = (1-\delta_{rs}/2)e_{rs}$. For the
multigraph case, see Appendix~\ref{app:multigraphs}. If we use the
shorthand $\{\theta\} = \{\{e_{rs}^l\},\{b_i\}\}$ for the model
parameters, the total likelihood of observing the layered graph is
\begin{equation}\label{eq:l_covariates}
  P(\{G_l\}|\{\theta\}) = P(G_c|\{\theta\})\prod_{r\leq s}\frac{\prod_lm_{rs}^l!}{m_{rs}!},
\end{equation}
where $P(G_c|\{\theta\}) = e^{-\mathcal{S}_t}$ is the likelihood of the
collapsed stochastic block model, where $\mathcal{S}_t$ is the
microcanonical entropy~\cite{peixoto_entropy_2012}. For instance, for
simple undirected graphs that are sparse (i.e. with $e_{rs} \ll
n_rn_s$), we have~\cite{peixoto_entropy_2012}
\begin{equation}
  \mathcal{S}_t \approx E - \frac{1}{2} \sum_{rs}e_{rs}\ln\frac{e_{rs}}{n_rn_s}.
\end{equation}

Here we are free to replace the traditional SBM by any other flavor,
which amounts simply to a different likelihood in the first term of
Eq.~\ref{eq:l_covariates}. The traditional SBM considered above imposes
that all nodes belonging to the same group will receive the same number
of edges on average, with little variation. An important alternative to
this is the degree-corrected stochastic block model
(DCSBM)~\cite{karrer_stochastic_2011}, that includes as additional model
parameters the degree sequence of the network, $\{k_i\}$. As argued in
Ref.~\cite{karrer_stochastic_2011}, and supported by an empirical model
selection analysis in Ref.~\cite{peixoto_model_2015}, this version is
often a better model for many (collapsed) networks that feature
significant degree variability. However, in this version with edge
covarites, only the degrees of the \emph{collapsed graph} are
constrained, and thus the edges incident on a specific node will be
distributed randomly among the layers independently of its
degree. Hence, for networks generated in this manner, nodes with a large
collapsed degree will also tend to possess uniformly larger degrees in
all layers, when compared to other nodes of the same group with a lower
collapsed degree. In other words, this model does not allow for degree
variability \emph{across layers}.

The complete likelihood of this model can be obtained in an entirely
analogous fashion, simply by augmenting the parameter set in
Eq.~\ref{eq:l_covariates} to include the collapsed degree sequence, i.e.
$\{\theta\} = \{\{e_{rs}^l\},\{b_i\}, \{k_i\}\}$, and using the
likelihood of the degree-corrected model~\cite{peixoto_entropy_2012}.

Other useful variations are SBMs with mixed memberships
(e.g.~\cite{airoldi_mixed_2008, ball_efficient_2011,
peixoto_model_2015}), in which nodes are allowed to belong to more than
one group. Here we use the formulation of
Ref.~\cite{peixoto_model_2015}, where we need to replace the node
partitions above by overlapping partitions, $\{\vec{b}_i\}$, where
$\vec{b}_i$ determines the mixture of node $i$, with $b_i^r \in \{0,
1\}$ specifying whether node $i$ belongs to group $r$, so that
$\{\theta\} = \{\{\vec{b}_i\}, \{e_{rs}\}\}$. Likewise, for the
degree-corrected version, we need to specify the (collapsed) labeled
degree sequence $\{\vec{k}_i\}$, where $k_i^r$ is the degree of node $i$
of type $r$, leading to $\{\theta\} = \{\{\vec{b}_i\}, \{e_{rs}\},
\{\vec{k}_i\}\}$. In both cases we simply replace the likelihood in
Eq.~\ref{eq:l_covariates} by the ones described in
Ref.~\cite{peixoto_model_2015}.

\subsection{SBM with independent layers}

Alternatively, we may generate each layer as an independent SBM,
constrained only by the fact that the group memberships of the nodes are
the same across all layers (although this can be relaxed in the
overlapping version, as discussed below). Furthermore, we allow nodes to
belong only to a subset of the layers, by including a $N\times C$ layer
membership matrix $\{z_{il}\}$, where each binary entry $z_{il} \in [0,
1]$ determines whether node $i$ belongs to layer $l$. If a node does not
belong to a given layer, it is forbidden to receive edges of that type.

Using the shorthand $\{\{\theta\}_l\} = \{\{e^l_{rs}\}\}$ and $\{\phi\}
= \{b_i\}$, the likelihood of the resulting layered block model is
simply
\begin{equation}\label{eq:l_layers}
  P(\{G_l\}|\{\{\theta\}_l\}, \{\phi\},\{z_{il}\}) = \prod_l P(G_l|\{\theta\}_l, \{\phi\}),
\end{equation}
with $P(G_l|\{\theta\}_l,\{\phi\})$ being the likelihood of the
traditional stochastic block model as before, where $G_l$ is the
subgraph containing only the edges of layer $l$ and the nodes specified
by $\{z_{il}\}$.

Like with the edge covariates model, here we are also free to replace
the traditional SBM by any other flavor, which amounts simply to
different likelihoods in the product of Eq.~\ref{eq:l_layers}. However,
differently from the SBM with edge covariates, if we wish to include
degree correction, we need to specify the \emph{layer-specific} degree
sequence $\{k^l_i\}$, where $k^l_i = \sum_jA^l_{ij}$ is the degree of
node $i$ in layer $l$, so that $\{\{\theta\}_l\} = \{\{e^l_{rs}\},
\{k^l_i\}\}$. Therefore, unlike the previous case, this model allows for
degree variability across different layers, i.e. a node with a large
degree in one layer, may possess very low degree in another. Note that
given the layer-specific degree sequence, we do not need to distinguish
between nodes that belong or not to a layer, since a node with a
layer-specific degree equal to zero will inherently not receive any edge
in that layer. Therefore the parameters $\{k^l_i\}$ replace the
parameters $\{z_{ij}\}$, which are removed from Eq.~\ref{eq:l_layers} in
this case.

We again may wish to use mixed-membership models in each layer, by using
overlapping partitions as parameters, i.e. $\{\phi\} =
\{\vec{b}_i\}$. For the degree-corrected version, we need to specify the
labeled degree sequence \emph{at each layer}, $\{\vec{k}_i\}_l$, where
${k_i^r}_l$ is the degree of node $i$ of type $r$ in layer $l$,
i.e. $\{\{\theta\}_l\} = \{\{e^l_{rs}\}, \{\vec{k}_i\}_l\}$. We may view
the labeled degree sequence inside each layer as a weighted membership
to each group. Since these ``weights'' may change across the layers
(even becoming zero), this corresponds to a generalization that allows
the memberships to change arbitrarily between the layers (despite the
fact that the overall, unweighted group mixtures $\{\vec{b}_i\}$ are
constant across the layers). This is a particularly useful property for
temporal networks, that allows group membership to change in time, as
discussed in more detail in Sec.~\ref{sec:time}.

\subsection{Equivalence between models}

The ``independent layers'' and ``edge covariates'' models are equivalent
in some situations, and different in others. In particular, in the
non-degree-corrected case described above, if all nodes belong to all
layers, both models generate the same networks asymptotically with the
same probability. This can be seen by employing Stirling's approximation
$\ln e^l_{rs}! \approx e^l_{rs}\ln{e^l_{rs}} - e^l_{rs}$ in
Eq.~\ref{eq:l_covariates}, which makes it identical to
Eq.~\ref{eq:l_layers}. Hence, as long as the edge counts in each layer
are sufficiently large, these models are fully equivalent. However, if
nodes belong only to specific subset of the layers, these models are not
equivalent. In this case, only the model with independent layers will
take the heterogeneous layer memberships into account, and hence it
should be preferred. Since we assume that the layer memberships are
known \emph{a priori} there is no reason to employ the ``edge
covariates'' non-degree-corrected model, since the ``independent
layers'' model will always provide an equal or better
description asymptotically\footnote{This may change if the layers are not entirely
known, and need to be determined, as in the case with real-valued
covariates in Sec.~\ref{sec:real}.}.

The situation is different for the degree-corrected models. Strictly,
both model versions are not equivalent, since the layered version allows
for degree variability across layers, whereas the covariate version does
not. Hence, there are networks generated by the layered model that
cannot be generated (or only with a vanishing probability) by the edge
covariates model. The opposite, however, is not true: A layered network
generated by the covariate version can always be sampled with the
independent layers version given an appropriate parameter choice.

Since the SBM with independent layers version always encapsulates the
edge covariate version, one might be tempted to prefer it
systematically. However, one needs to realize that the layered version
requires more parameters than the covariates version, either via the
layer membership matrix $\{z_{il}\}$ or the layer-specific degree
sequence $\{k^l_i\}$. Similar comparisons can be made between specific
flavors of both models (e.g. with overlapping groups or degree
correction). Because of the increased number of degrees of freedom in
the model specification, we risk overfitting the data by always choosing
the most constrained model. We discuss exactly how this choice between
models should be done in the next section.

\section{Selecting the most appropriate model}\label{sec:selection}

The proper way to select between alternatives is to perform model
selection based on statistical significance, and opt for the more
complicated model only if there is sufficient evidence available in the
data to compensate the larger number of parameters. Formulated in a
Bayesian setting, as proposed in Ref.~\cite{peixoto_model_2015}, this
selection procedure amounts to finding the model that maximizes the
posterior likelihood
\begin{equation}
  P(\{\theta\} | \{G_l\}) = \frac{P(\{G_l\}|\{\theta\})P(\{\theta\})}{P(\{G_l\})},
\end{equation}
where $\{\theta\}$ is a shorthand for the entire set of model parameters
(e.g. for the non-degree-corrected SBM with edge covariates we have
$\{\theta\} = \{\{b_i\},\{e_{rs}^l\}\}$), $P(\{\theta\})$ is the prior
probability on the parameters, and $P(\{G_l\})$ is a normalization
constant. Since in our context we are dealing with discrete parameters,
we can write $P(\{\theta\}) = e^{-\mathcal{L}(\{\theta\})}$, where
$\mathcal{L}(\{\theta\})$ is the microcanonical entropy of the parameter
ensemble. Therefore, we have that $-\ln P(\{\theta\} | \{G_l\}) = \Sigma
+ \ln P(\{G_l\})$ with $\Sigma = \mathcal{S}(\{G_l\}) +
\mathcal{L}(\{\theta\})$ being the description length of the
data~\cite{grunwald_minimum_2007, rissanen_information_2010,
rosvall_information-theoretic_2007}. Hence this approach amounts to
finding the model that most compresses the observed data, i.e. the one
with the minimum description length, since to maximize $P(\{\theta\} |
\{G_l\})$ is equivalent to minimize
$\Sigma$~\cite{peixoto_parsimonious_2013, peixoto_hierarchical_2014,
  peixoto_model_2015}.

Here we observe that since the prior probabilities are nonparametric,
the whole procedure also becomes parameter-free, and hence no \emph{ad
hoc} choices are required \emph{a priori}. In particular for the SBM
variants considered in this work, the partition of the nodes, degree of
overlap, the number of groups and the hierarchical structure are
obtained in entirely nonparametric fashion.

\subsection{Choice of priors}

In order to compute $P(\{\theta\})$, we need to describe generative
processes for the parameter themselves. This means that for the model
variants above we need to specify a generative process for the partition
into $B$ groups $\{b_i\}$, the layer membership matrix $\{z_{il}\}$, the
collapsed (or layer-specific) degree-sequence $\{k_i\}$ (or $\{k_i^l\}$),
and the layered edge counts $\{e_{rs}^l\}$. (In the overlapping case, we
need to do the same for the overlapping partition and labeled degree
sequences, which we show in Appendix~\ref{app:overlap}.)

Choosing prior probabilities is a subtle issue, since it depends on
\emph{a priori} assumptions about the data, which usually depends on
context, and often requires domain-specific knowledge. In general
situations, a prudent approach is to choose uninformative priors, which
do not bias the estimation. Here we will take the systematic approach of
choosing a nested sequence of priors and hyperpriors, so that an
uninformative prior is chosen only at the topmost
level~\cite{peixoto_hierarchical_2014, peixoto_model_2015}. This
approach is intended to minimize the sensitivity of the choice of
priors, and accordingly provide a shorter description length in the
majority of cases.

To generate the partition into groups, we use the process described in
detail in Refs.~\cite{peixoto_hierarchical_2014, peixoto_model_2015},
that corresponds to a multilevel Bayesian process, where the
distribution of group sizes $\{n_r\}$ (where $n_r$ is the number of
nodes in group $r$) is first uniformly sampled from the set of all
allowed possibilities, and the partition is distributed uniformly,
conditioned of the observed size distribution, yielding a description
length $\mathcal{L}_p=-\ln P(\{b_i\})$ given by
\begin{align}\label{eq:lp}
  \mathcal{L}_p = \ln{\textstyle\multiset{B}{N}} + \ln N! - \sum_r\ln n_r!,
\end{align}
where $\multiset{n}{m} = {n + m - 1 \choose m}$ is the total number of
$m$-combinations with repetitions from a set of size $n$.

For the independent layers model without degree correction, we need to
specify the node memberships to each layer. For this, we use the process
described in detail in Ref.~\cite{peixoto_model_2015} to generate
overlapping partitions. We represent each line in the $\{z_{il}\}$
matrix as a mixture vector $\vec{z}_i$ with $C$ binary entries. We
formulate a multilevel Bayesian process, where the distribution of
mixture sizes $\{n_d\}$ (where $d_i=\sum_lz^l_i$ is the mixture size of
node $i$, and $n_d$ is the number of nodes with $d_i=d$) is generated
from all possibilities with uniform probability, and the local values of
$d_i$ are sampled from this distribution. The mixture distribution
$\{n_{\vec{z}}\}$ (where $n_{\vec{z}}$ is the number of nodes belonging
to mixture $\vec{z}$) is also sampled from the set of possible choices
with uniform probability, conditioned of the local mixture sizes
$\{d_i\}$, and finally the individual mixtures $\{\vec{z}_i\}$
themselves are sampled from this distribution. This yields a description
length $\mathcal{L}_z=-\ln P(\{\vec{z}_i\})$ given
by~\cite{peixoto_model_2015}
\begin{align}\label{eq:lz}
  \mathcal{L}_z = \ln{\textstyle\multiset{C}{N}} + \sum_d \ln {\textstyle\multiset{{C \choose d}}{n_d}} + \ln N! - \sum_{\vec{z}}\ln n_{\vec{z}}!.
\end{align}

The \emph{collapsed} degree sequence can be generated with a similar
Bayesian process, described also in Ref.~\cite{peixoto_model_2015}, that
yields a description length $\mathcal{L}_{\kappa} = -\ln
P(\{k_i\})$ given by
\begin{equation}\label{eq:lkappa_D1}
  \mathcal{L}_{\kappa} = \sum_r\min\left(\mathcal{L}^{(1)}_r, \mathcal{L}^{(2)}_r\right),
\end{equation}
with
\begin{align}
  \mathcal{L}^{(1)}_r &= \ln{\multiset{n_r}{e_r}}, \\
  \mathcal{L}^{(2)}_r &= \ln\Xi_r + \ln n_r! - \sum_k \ln n^r_k!,
\end{align}
and $\ln\Xi_r \approx 2\sqrt{\zeta(2)e_r}$.

For layered networks, we need a generative process for the
layer-specific degree sequence, $\{k_i^l\}$. Although one could in
principle construct nonparametric distributions that incorporate
arbitrary correlations among the degree sequences of all layers, the
dimension of such distributions is likely to exceed the evidence
available in typical data as the number of layers increases. Therefore,
here we take the simpler route and assume independent distributions at
each layer, so that the description length $\mathcal{L}_{\varkappa} =
-\ln P(\{k_i^l\})$ becomes simply
\begin{equation}\label{eq:lkappa}
  \mathcal{L}_{\varkappa} = \sum_l\mathcal{L}_{\kappa}(\{k_i\}^l),
\end{equation}
where $\{k_i\}^l$ should be understood as the collapsed degree sequence
of the graph containing only the edges belonging to layer $l$.

Finally, to generate the edge counts $\{e^l_{rs}\}$, we note that they
can be viewed as the adjacency matrix of a layered multigraph with $B$
nodes~\cite{peixoto_parsimonious_2013}. Therefore, we may use the
stochastic blockmodel itself to generate it, either with independent
layers or edge covariates. Since these models have their own edge count
parameters, this forms a nested sequence of SBMs, encapsulating the
multilevel hierarchical structure of the network, in a fully
nonparametric fashion, yielding a description length as described in
Ref.~\cite{peixoto_hierarchical_2014},
\begin{equation}\label{eq}
  \mathcal{L}_e = \sum_{h=1}^LS_m(\{e^l_{rs}\}^h, \{n_r\}^h) + \sum_{h=1}^{L-1}\mathcal{L}^h_p,
\end{equation}
where $S_m(\{e^l_{rs}\}^h, \{n_r\}^h)$ is the appropriate entropy of the
layered SBM in hierarchical level $h$, and $\mathcal{L}^h_p$ is the
description length of the corresponding node partition.

At the top of the hierarchy we have the remaining parameters $\{E_l\}$,
denoting the number of edges in each layers. For completeness, they can
be easily generated by including an uniform prior
$P(\{E_l\})=\multiset{L}{E}^{-1}$, however this only adds an overall
constant to the description length, which is not relevant to any
comparisons made in this paper.

To summarize, using the shorthand $\{\theta\}$ for the entire set of
parameters, we have for each given model (i.e. edge covariates and
independent layers, with any optional combination of degree correction
and group overlap) an overall description length
\begin{equation}\label{eq:sigma}
  \Sigma = \mathcal{S}(\{\theta\}) + \sum_{\theta}\mathcal{L}_\theta,
\end{equation}
where $\mathcal{S}(\{\theta\})$ is the appropriate SBM entropy, and
$\mathcal{L}_\theta$ is the description length of a specific parameter
ensemble, chosen from Eqs.~\ref{eq:lp} to~\ref{eq:lkappa} (and
Eqs.~\ref{eq:o_lp} to~\ref{eq:o_lkappa}), as appropriate.

\subsection{Confidence levels}

As described above, selecting the model with the smallest description
length $\Sigma$ is the appropriate manner of balancing model complexity
and goodness of fit. However, often we desire a more refined approach
where the alternative model can be accepted or rejected with a degree of
confidence, in a nonparametric fashion. This can be achieved, as
proposed in Ref.~\cite{peixoto_model_2015}, by inspecting the posterior
odds ratio~\cite{jaynes_probability_2003},
\begin{align}\label{eq:lambda}
  \Lambda &= \frac{P(\{\theta\}_a|\{G_l\},\mathcal{H}_a)P(\mathcal{H}_a)}{P(\{\theta\}_b|\{G_l\},\mathcal{H}_b)P(\mathcal{H}_b)}\\
  & = \exp\left(-\Delta\Sigma\right)\frac{P(\mathcal{H}_a)}{P(\mathcal{H}_b)},
\end{align}
where $P(\{\theta\}|\{G_l\},\mathcal{H})$ is the posterior according to
hypothesis $\mathcal{H}$ (i.e. a specific model class), $P(\mathcal{H})$
is any prior belief for hypothesis $\mathcal{H}$, and $\Delta\Sigma =
\Sigma_a - \Sigma_b$ is the difference in description length between
both hypotheses. For $\Lambda < 1$ we have that $\mathcal{H}_a$ is
rejected over $\mathcal{H}_b$ with a confidence that increases as
$\Lambda$ decreases. Often the values of $\Lambda$ are divided in
subjective intervals of evidence strength~\cite{jeffreys_theory_1998},
as a convention with $\Lambda = 1/100$ being considered the plausibility
threshold, below which $\mathcal{H}_a$ is decisively rejected in favor
of $\mathcal{H}_b$, and with $\Lambda \in [1/3, 1]$ being considered
only a negligible difference between both models. In the case where
there is no preference for either model, $P(\mathcal{H}_a) =
P(\mathcal{H}_b)$, the value of $\Lambda$ is called the Bayes
factor~\cite{jeffreys_theory_1998}, which has the same
interpretation. In the following, we will always assume
$P(\mathcal{H}_a) = P(\mathcal{H}_b)$, and impose $\Lambda \leq 1$, by
always putting the preferred hypothesis in the denominator of
Eq.~\ref{eq:lambda}.

\subsection{Inference algorithm}

The description length of a given flavor of the SBM given by
Eq.~\ref{eq:sigma} is an objective function that needs to be minimized
with some appropriate algorithm. The only known algorithm that is
guaranteed to find the global minimum is the exhaustive computation of
the description length for every possible hierarchical partition of the
network, which is unfeasible in any practical scenario with networks
with more than a few nodes and edges. Therefore, we must resort to
approximate methods. Here we employ the multilevel MCMC algorithm
described in Ref.~\cite{peixoto_efficient_2014}, together with the
hierarchical generalization presented in
Ref.~\cite{peixoto_hierarchical_2014}, and the extension to overlapping
groups presented in Ref.~\cite{peixoto_model_2015}. The advantage of
these algorithms is their good typical running times, and their capacity
to overcome metastable states by performing agglomerative
moves~\footnote{We note that in principle other algorithms such as
belief propagation~\cite{decelle_asymptotic_2011} and spectral
clustering~\cite{krzakala_spectral_2013} could be used as well, provided
their are suitably adapted to the nonparametric likelihoods considered
here.}.  The division of the network into layers does not alter these
algorithms in any significant way, other than a straightforward
book-keeping of the layer membership of each edge. In particular, by
using appropriate sparse data structures that do not change in size if
the number of layers is increased, the division into layers does not
alter significantly the typical running times of the algorithms, which
remain $O(N\ln^2N)$ in their greedy versions, independent of the number
of groups $B$ and layers $C$, and hence are applicable to reasonably
large networks. An efficient C++ implementation of these algorithms is
freely available as part of the {\texttt{graph-tool}} Python
library~\cite{peixoto_graph-tool_2014}
at~\url{http://graph-tool.skewed.de}.

\section{When are layers informative?}\label{sec:info}

\begin{figure}
  \begin{tabular}{cc}
     \rotatebox{-8}{\includegraphics[width=0.4\columnwidth]{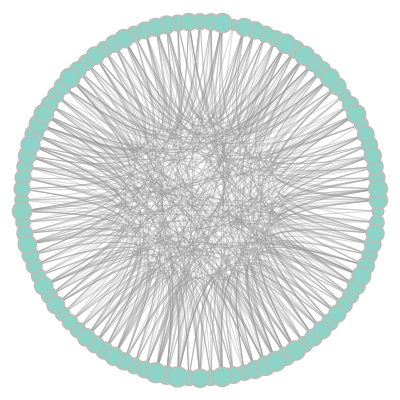}} & \rotatebox{-8}{\includegraphics[width=0.4\columnwidth]{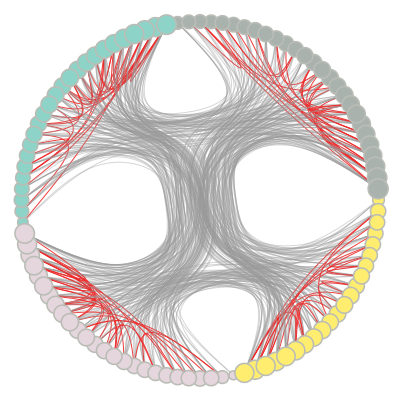}} \\
    (a) & (b)
  \end{tabular} \caption{(Color online) Artificial network example
  containing an informative layered structure. (a) The collapsed graph
  possesses no discernible structure, i.e. it corresponds to a fully
  random graph. (b) When the division of edges into two layers [grey and
  red (light grey)] is taken into account, a four-group structure is
  revealed.\label{fig:informative_layers}}
\end{figure}

Layers are informative of the network structure if their incorporation
into the model yields a more detailed description of the data, when
compared to a model that is only based on the collapsed structure of the
network.  An illustration of an informative layered structure is shown
in Fig.~\ref{fig:informative_layers}. In this example, an artificial
network composed of two layers is constructed. The collapsed graph
corresponds to a fully random network, however the division of the edges
into layers is such that four fully assortative groups exist in one of
the layers. Clearly, the layered division yields structural information
that is not discernible in the collapsed graph. This implies that, in
more general cases, omitting such information on the edges could
potentially significantly obscure structure present in the
data~\cite{kivela_multilayer_2014, boccaletti_structure_2014}.

\begin{figure}
  \begin{center}
    \hspace{1.5em}\includegraphics{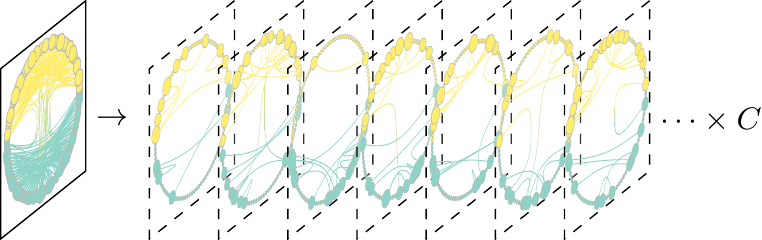}
    \vspace{1em}
    \includegraphics[width=.8\columnwidth]{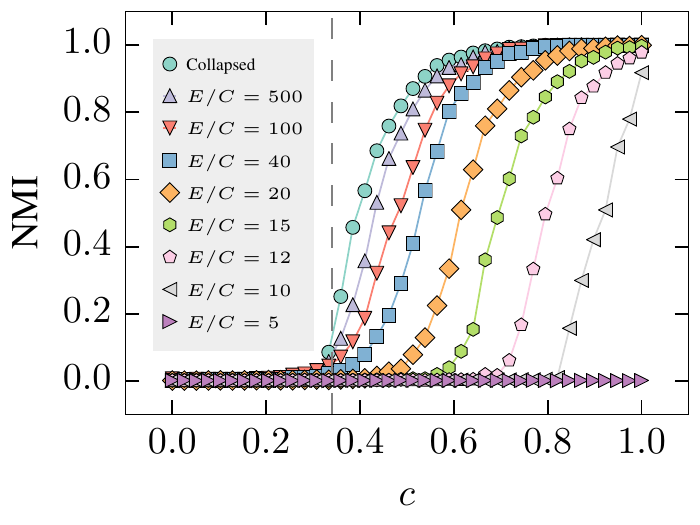}
    \includegraphics[width=.8\columnwidth]{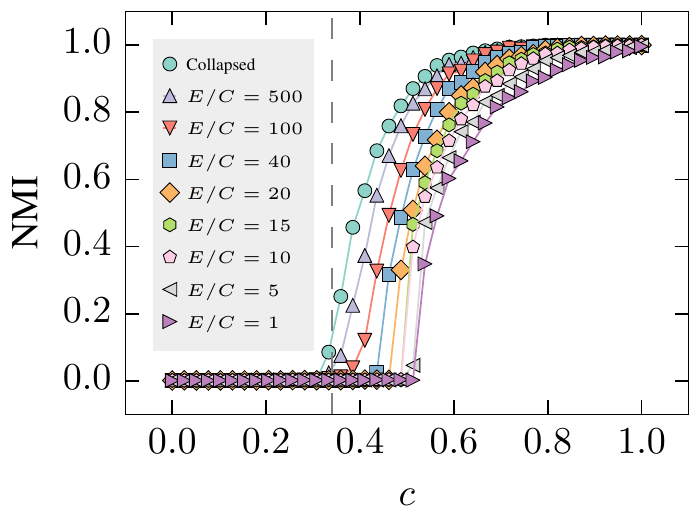}
  \end{center}
  \vspace{-2em}

  \caption{(Color online) An excessive number of layers can obscure
  network structure. \emph{Top}: A collapsed two-group structure is
  generated, and the edges are randomly distributed in $C$
  layers. \emph{Middle and Bottom:} As the number of edges per layer
  $E/C$ diminishes, the structure inside each layer becomes increasingly
  sparse, and the overall quality of the inference worsens. The middle
  panel shows the normalized mutual information (NMI) between the
  planted and inferred partitions, using the SBM with independent
  layers, for a network of $N=10^4$ nodes and average degree
  $\avg{k}=2E/N=14$ as a function of the mixing parameter $c$, as
  described in the text. The bottom panel is the same as the middle one,
  but using the SBM with edge covariates. In both cases the vertical
  lines mark the detectability transition point for the collapsed SBM,
  $c^* = 1/B +
  (B-1)/(B\sqrt{\avg{k}})$~\cite{decelle_inference_2011}.\label{fig:noninformative_layers}}
\end{figure}

However, it is important to realize that the opposite is also true: If
the edge distribution into layers is uncorrelated with the group
divisions, it can also obscure structural information which would
otherwise be revealed if the layer information were to be ignored. This
happens because increasing the number of layers in the model also
increases its effective dimension. If the total size and density of the
network remains constant as the number of layers increases (and hence
the effective dimension of the model), the available data become
increasingly sparse, which reduces the inference precision, since it
becomes increasingly difficult to distinguish signal from noise. An
example of this is shown in Fig.~\ref{fig:noninformative_layers},
corresponding to a collapsed $B=2$ assortative SBM with equal-sized
groups and edge counts given by $e_{rs}=2E[\delta_{rs}c/B
+(1-\delta_{rs})(1-c)/B(B-1)]]$, with $c\in [0, 1]$ being a mixing
parameter, where the edges are distributed randomly in $C$ layers. As
$C$ increases, both model variants (edge covariates, and independent
layers) display increasing degradation when inference is performed, with
the detectability transition~\cite{decelle_inference_2011} shifting to
higher values of $c$. For the SBM with independent layers, the
transition shifts to $c^*\to 1$ as $C\to E$, and in this limit no
information at all on the graph structure can be inferred.  The version
with edge covariates displays a relatively superior performance, with
the transition remaining at $c^*<1$ for $C\to E$, since it is
conditioned on the collapsed graph. Nevertheless, even in this case the
degradation caused by increasing $C$ is very noticeable.

\begin{figure}
  \centering
  \begin{tabular}{cc}
  \includegraphics[width=.45\columnwidth]{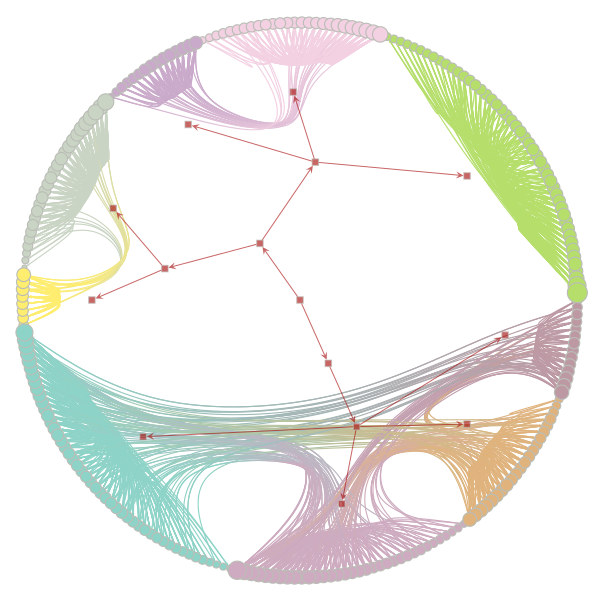} &
  \includegraphics{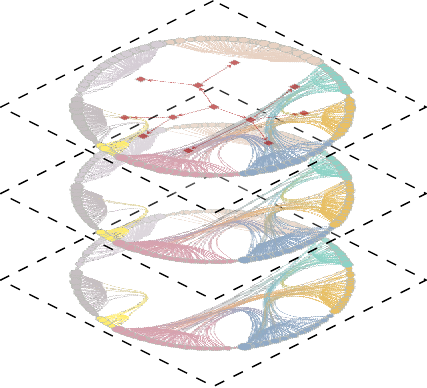}\\
  (a) $\Lambda=1$ & (b) $\log_{10}\Lambda \approx -51$
  \end{tabular} \caption{(Color online) \label{fig:physicians} Two generative models
  for a layered social network of
  physicians~\cite{coleman_diffusion_1957}. (a) Inferred DCSBM for the
  collapsed network, with the edges assumed to be randomly distributed
  among the layers. (b) Inferred DCSBM with edge covariates, where each
  layer corresponds to one type of acquaintance. Below each figure is
  shown the posterior odds ratio $\Lambda$, relative to preferred model
  (a). The circular layout with edge
  bundling~\cite{holten_hierarchical_2006} represents the inferred node
  hierarchy (indicated also by the red nodes and edges), as explained in
  the text (see also Ref.~\cite{peixoto_hierarchical_2014}).}
\end{figure}

Because of this problem, it is important to consider if we indeed need
the layered structure to describe the large-scale structure of a
network, or if it needs to be coarse-grained or even discarded. This can
be done by considering a null model where the edges are distributed
among the layers in a manner that is entirely independent of the group
structure, and is parametrized only by the total number of edges in each
layer, $\{E_l\}$. Let us use the shorthand $\{\theta\}$ for the possible
set of parameters of a collapsed SBM. This null model has a likelihood
given simply by
\begin{equation}\label{eq:null}
  P(\{G_l\}|\{\theta\},\{E_l\}) =
  P(G_c|\{\theta\})\times \frac{\prod_lE_l!}{E!}
\end{equation}
where the first term is the likelihood of the collapsed SBM and the
second accounts for the random distribution of edges across the layers
(the above equation is valid only for simple graphs; for multigraphs see
Appendix~\ref{app:multigraphs}). The full posterior and its
corresponding description length are computed just as before, by
including the priors for $\{e_{rs}\}$, $\{b_i\}$, $\{\vec{b}\}$,
$\{k_i\}$ and $\{\vec{k}_i\}$. We can then compare the description
length of this null model with any of the other layered variants, and
decide if there is enough evidence to justify the incorporation of
layers that are correlated with the group structure.

As a concrete example, here we consider an empirical social network of
$N=241$ physicians, collected during a
survey~\cite{coleman_diffusion_1957}. Participants were asked which
other physicians they would contact in hypothetical situations. The
questions asked were: 1. ``When you need information or advice about
questions of therapy where do you usually turn?'', 2. ``And who are the
three or four physicians with whom you most often find yourself
discussing cases or therapy in the course of an ordinary week
-- last week for instance?'', 3. ``Would you tell me the first names of
your three friends whom you see most often socially?''. The answers to
each question represent edges in one specific layer of a directed
network. If one applies the DCSBM to the collapsed graph (which provides
the best fit among the alternatives), it yields a division into $B=9$
groups, as shown in the left panel of Fig.~\ref{fig:physicians},
including also a division into three disconnected components
(corresponding to different cities). Between the layered SBM versions,
the model with edge covariates that turns out to be a better fit to the
data (i.e. yields a lower description length) and divides the network
into $B=8$ groups, as shown in the left panel of
Fig.~\ref{fig:physicians}. When inspecting the edge counts visually, one
does not notice any significant difference between the patterns in each
layer. Indeed, when comparing the description lengths between the null
model with random layers above and the SBM with edge covariates, we find
that the latter is strongly rejected with a posterior odds ratio
$\Lambda \approx 10^{-51}$. Therefore, there is no noticeable evidence
in the data to support any correlation of layer divisions with the
large-scale structure present in the graph. This suggests that the
important descriptors of this social network are mainly the overall
acquaintances among physicians, not their precise types (at least as
measured by the survey questions).

\begin{figure*}
  \centering
  \begin{tabular}{cc}
  \includegraphics[width=.4\textwidth]{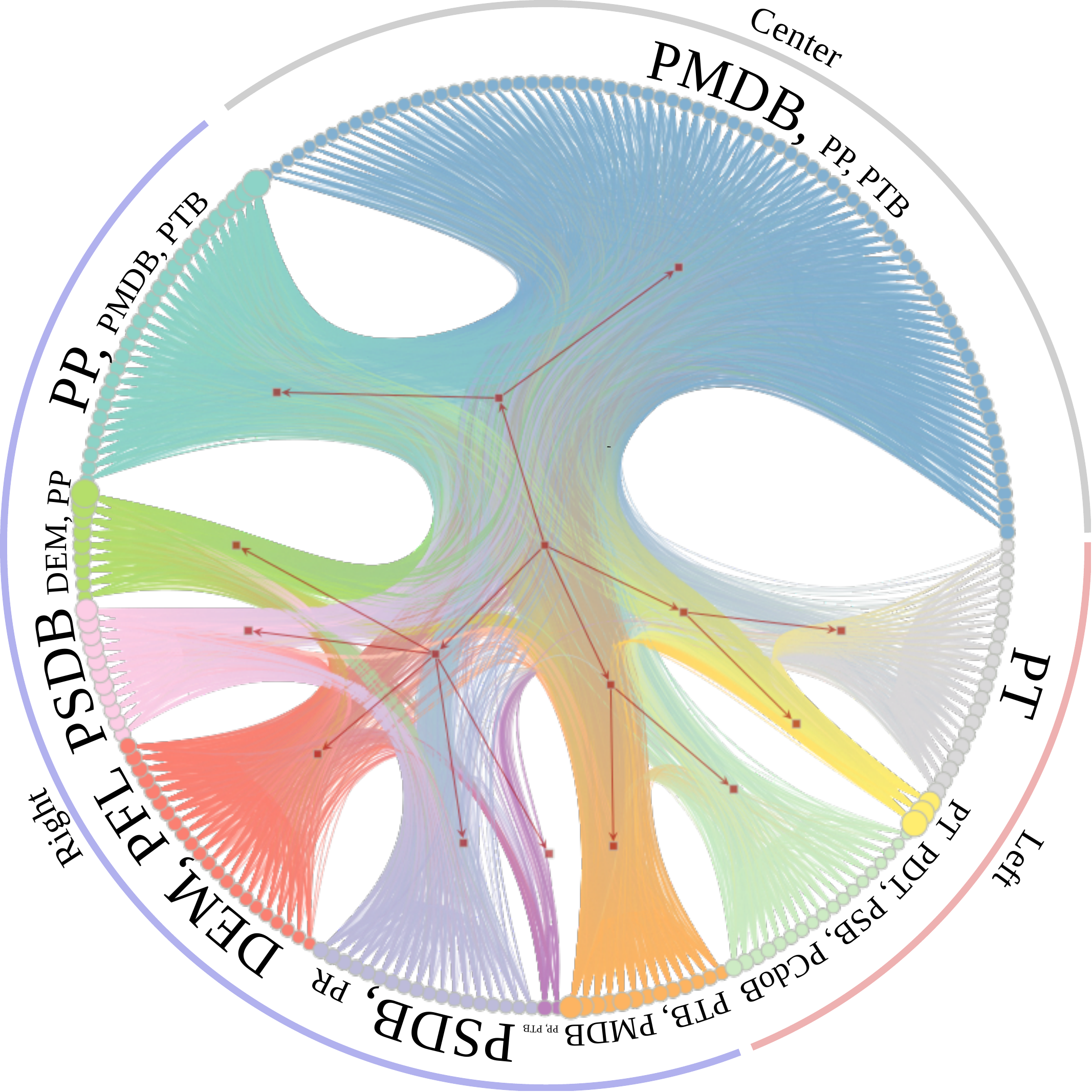} &
  \includegraphics{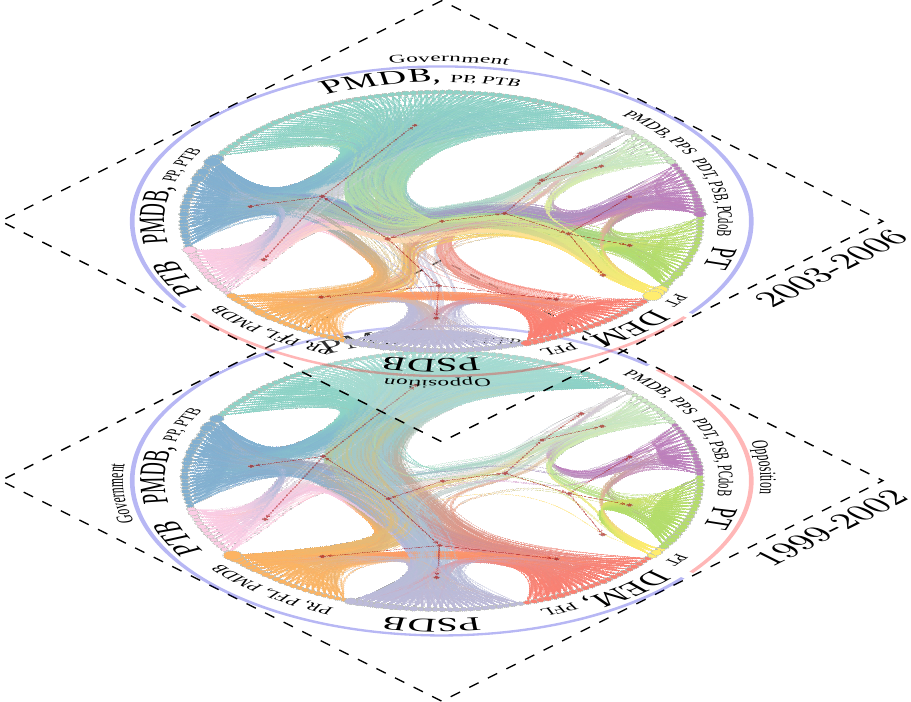}\\
  (a) $\log_{10}\Lambda \approx -111$ & (b) $\Lambda=1$
  \end{tabular} \caption{(Color online) \label{fig:camara} Network of vote correlations
  among federal deputies of the Brazilian national congress during two
  consecutive four-year terms, $1999-2002$ and $2003-2006$. (a) DCSBM
  fit for the collapsed network obtained by merging both terms,
  corresponding to a null model where the edges are randomly distributed
  between the layers. The group labels correspond to the predominant
  parties inside each group, determined after the inference had been
  performed (the size of the label indicates the proportion of each
  party inside the group). (b) DCSBM with independent layers for the
  network divided into two terms. In both cases is shown the posterior
  odds ratio $\Lambda$ relative to the best model [in this case
  (b)]. The layout is the same as in Fig.~\ref{fig:physicians}.}
\end{figure*}

We now turn to another example, where informative layered structure can
be detected. We consider the vote correlation network of federal
deputies in the Brazilian national congress. Based on public data
containing the votes of all deputies in all chamber sessions across many
years\footnote{Available at \url{http://www.camara.gov.br/}.}, we
obtained the correlation matrix between all deputies. We constructed a
network by connecting an edge from a deputy to other 10 deputies with
which she is most correlated in the considered period\footnote{We
experimented with other threshold values, and obtained similar
results.}. We then separated the network in two layers, corresponding to
two consecutive four-year terms, $1999-2002$ and $2003-2006$. Deputies
not present during the whole period were removed from the network,
yielding a network with $N=224$ nodes and $E=7247$ edges in total.  When
fitting the DCSBM for the collapsed network (which is again the best
model), we obtain the $B=11$ partition shown in the left panel of
Fig.~\ref{fig:camara}. It shows a hierarchical division that is largely
consistent with party and coalition lines, as well as positions in the
political spectrum (with a noticeable deviation being a group of
left-wing parties composed by PDT, PSB and PCdoB being grouped together
with center-right parties PTB and PMDB).  When incorporating the layers,
the best model fit is obtained by the DCSBM with independent layers,
which yields a $B=11$ division mostly compatible with (but not fully
identical to) the collapsed network, although with a different
hierarchical structure, as can be seen in the right panel of
Fig.~\ref{fig:camara}. However, the layered representation of this
network reveals a major coalition change between the two terms,
consistent with the shift of power that occurred with the election of a
new president belonging to the previous main opposition party: In the
$1999-2002$ term we see a clear division into a government and
opposition groups (as captured in the topmost level of the hierarchy),
with most edges existing between groups of the same camp, corresponding
to a right-wing/center government led by the PSDB, PMDB, PFL, DEM and PP
parties, and a left-wing opposition composed mostly by PT, PDT, PSB and
PCdoB. After $2002$, we observe a shifted coalition landscape, with a
left-wing/center government predominantly formed by PT, PMDB, PDT, PSB
and PCdoB, and an opposition led by PSDB, PFL, DEM and PP. Because of
this noticeable change in the large-scale network structure --- that is
completely erased in the collapsed network --- the null model with
random layers ends up being forcefully rejected with $\Lambda \approx
10^{-111}$, meaning that the layered structure is very informative on
the network structure.

In the above examples we made a comparison between the layered model and
a null model with fully random layers. In some scenarios we might be
interested in a more nuanced approach, where the layers are
coarse-grained with a more appropriate level of granularity. This can be
done by merging some of the layers into bins, such that inside each bin
the layer membership of the edges is distributed regardless of the
group structure. Let $\ell$ specify a set of layers that were merged in
one specific bin, and $\{\theta\}_{\{\ell\}}$ be a shorthand
for the possible set of parameters of a layered SBM $\{G_\ell\}$ (with
independent layers or edge covariates) where each bin $\ell$ corresponds
to an individual layer. The likelihood of this model conditioned on a
specific bin set $\{\ell\}$ is is given by
\begin{equation}\label{eq:p_ell}
  P(\{G_l\}|\{\theta\}_{\{\ell\}},\{\ell\})  =
  P(\{G_\ell\}|\{\theta\}_{\{\ell\}})\times \prod_\ell\frac{\prod_{l\in\ell}E_{l}!}{E_\ell!},
\end{equation}
where $E_\ell = \sum_{l\in \ell}E_l$ is the number of edges in bin
$\ell$ (the above equation is valid only for simple graphs; See
Appendix~\ref{app:multigraphs} for the more general case with parallel
edges).  When considering the full posterior, we need to include the
priors for $\{\theta\}_{\{\ell\}}$ as before, but also for the binning
$\{\ell\}$ itself. If the layers can be grouped arbitrarily, we have
\begin{equation}
  P(\{\ell\}) = \frac{\prod_{\ell}n_{\ell}!}{C!}\times\multiset{M}{C}^{-1}
\end{equation}
where $n_{\ell}$ is the number of layers in bin $\ell$ and $M$ is the
total number of layer bins. If the layers are inherently ordered, and
thus can only be contiguously binned, this becomes instead simply
\begin{equation}\label{eq:lbins}
   P(\{\ell\}) =\multiset{M}{C}^{-1}.
\end{equation}
If we make $M=1$ we recover the original null model above.
Algorithmically, one can find the appropriate bins in a variety of
ways. A simple approach is to use agglomerative hierarchical clustering,
i.e. by putting at first each layer in its own bin, and subsequently
merging bins according to the reduction of the overall description
length. We explore this idea further in Sec~\ref{sec:real}, when dealing
with real-valued edge covariates.

\subsection{Layers as evidence for overlaps}

There is an important correspondence between layered networks and
overlapping structures of collapsed networks. Namely, the inference of
overlapping structures in collapsed graphs can to some extent be
interpreted as the inference of latent
layers~\cite{valles-catala_multilayer_2014} to which the edges belong,
where each (connected) group pair $(r,s)$ would correspond to a
different layer. Because of this correspondence, any a priori knowledge
of the division into layers can fundamentally alter the interpretation
of the data in situations where a nonoverlapping model would otherwise
be considered a better fit for the collapsed
network~\cite{peixoto_model_2015}.

\begin{figure}
  \centering

  \begin{tabular}{cc}
  \includegraphics[width=.49\columnwidth]{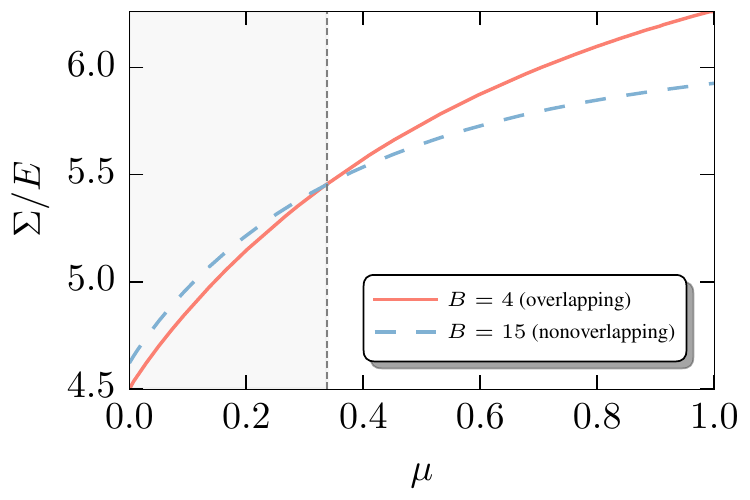}  &
  \includegraphics[width=.49\columnwidth]{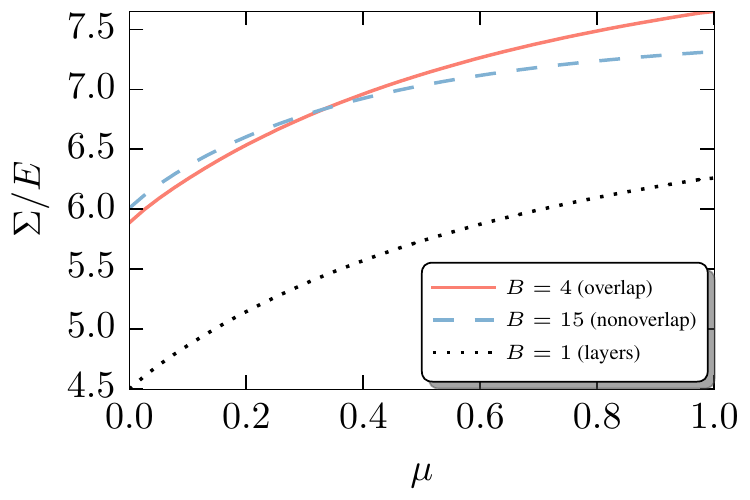}
  \end{tabular}

  \begin{tabular}{ccc}
  \includegraphics[width=.32\columnwidth]{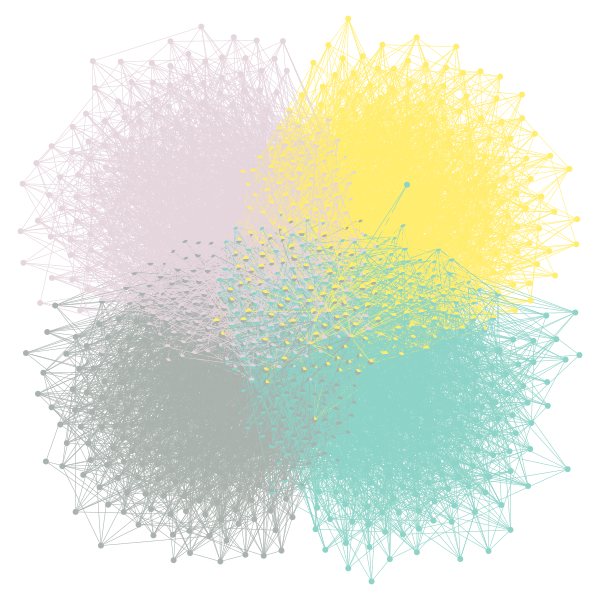}  &
  \includegraphics[width=.32\columnwidth]{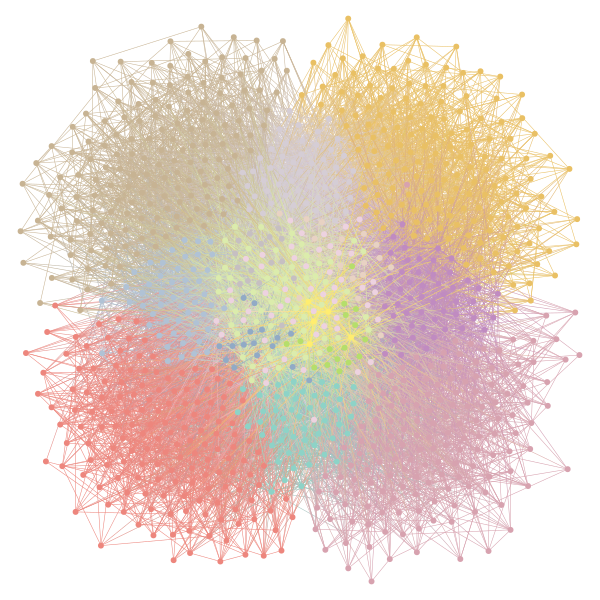} &
  \begin{minipage}[t]{.32\columnwidth}
  \includegraphics{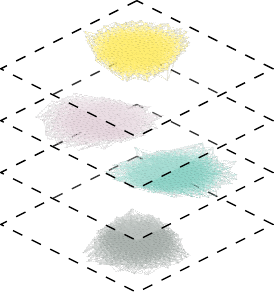}
  \end{minipage}
  \end{tabular}

  \caption{(Color online) \label{fig:evidence} \emph{Top left:} Description length per
  edge $\Sigma/E$ for the collapsed planted partition model described in
  the text as a function of the overlap parameter $\mu$, with $N=10^3$,
  $\avg{k}=2E/N=10$ and $B=4$ (illustrated in the lower left panel). The
  two curves show the description length of the planted overlapping
  model, and the equivalent non-overlapping model with $2^B-1$ groups
  (illustrated in the lower middle panel). Only for values of $\mu$
  below the intersection point the original overlapping model is
  preferred over the nonoverlapping one. \emph{Top right:} The same as
  in the top left, but with layer information included. The third curve
  corresponds to a $B=1$ model with $C=4$ independent layers
  (illustrated in the lower right panel), whereas the first two curves
  correspond to the same collapsed models as in the left panel, but with
  a random distribution of edges in the $C=4$ layers.  The model with
  independent layers is preferred over the alternatives in the entire
  parameter range.}
\end{figure}

This is better understood by considering the following generative
process as an example: A network is generated with $C$ layers, where in
each layer $E/C$ edges are randomly placed between the nodes that belong
to that layer. The layer membership mixtures are parameterized as
$n_{\vec{z}} \propto \prod_l\mu^{z_l}$, up to a normalization constant,
and with $\mu \in [0,1]$ controlling the degree of layer overlap: For
$\mu \to 0$ we obtain asymptotically nonoverlapping layers with $n_l =
N/B$ nodes at each layer $l$, and for $\mu=1$ all mixtures $\vec{z}$
have the same size. This process corresponds to a layered SBM with only
one group, $B=1$, and the aforementioned layer structure.  If we
consider only the collapsed graph, with the layer information removed,
the corresponding topology can be generated in two alternative ways:
1. An overlapping SBM with $B=C$ groups and mixtures
$\vec{b}_i=\vec{z}_i$, and edge counts $e_{rs} = 2E \delta{rs}/B$. 2. A
nonoverlapping SBM with each individual mixture as its own group,
indexed by $r_{\vec{b}}=\sum_{s=1}^Bb_s2^{s-1}\in [1, 2^C-1]$, resulting
in a total of $B=2^C- 1$ groups, and edge counts given by
\begin{equation}
  e_{r_{\vec{b}_1}r_{\vec{b}_2}} = \sum_{rs}b_1^rb_2^s\frac{e_{rs}}{n_rn_s}n_{r_{\vec{b}_1}}n_{r_{\vec{b}_2}}.
\end{equation}
The description length of the \emph{collapsed graph} generated with the
layered model is
\begin{equation}
  \Sigma_c = 2E - E\ln\frac{2EC}{N^2} + \mathcal{L}_z(\{n_{\vec{z}}(\mu)\}),
\end{equation}
which is in fact identical to the overlapping SBM, corresponding to
$C\to B$ and $n_{\vec{z}}(\mu) \to n_{\vec{b}}(\mu)$ in the above
equation. The nonoverlapping model, on the other hand, has a description
length given by
\begin{equation}
  \Sigma_c' = 2E - \frac{1}{2}\sum_{\vec{b}_1\vec{b}_2} e_{r_{\vec{b}_1}r_{\vec{b}_2}} \ln \frac{e_{r_{\vec{b}_1}r_{\vec{b}_2}}}{n_{r_{\vec{b}_1}} n_{r_{\vec{b}_2}}} + \mathcal{L}_p(\{n_{r_{\vec{b}}}(\mu)\}),
\end{equation}
where $\mathcal{L}_p(\{n_{r_{\vec{b}}(\mu)}\})$ corresponds to a
nonoverlapping partition of individual mixtures. As discussed in
Ref.~\cite{peixoto_model_2015}, we may have $\Sigma_t' < \Sigma_t$ if
the number of nodes at the intersections is sufficiently
large. Therefore the nonoverlapping model may indeed be considered the
most parsimonious of the three in that case, which is arguably
non-intuitive, since the overlapping SBM seems closer to the original
model. However, the situation changes when the observed data includes
the layer information on the edges. In this case, we must include the
random division of the edges into layers in the two collapsed models, by
adding, according to Eq.~\ref{eq:null}, the following term to the
description length:
\begin{equation}
  \ln E! - \sum_l \ln E_l! = \ln E! - C\ln E/C!.
\end{equation}
Because of this difference, the layered model with $B=1$ becomes always
the preferred choice (see Fig.~\ref{fig:evidence}). Therefore, when edge
information is available, it can significantly change which model is
preferred, and tip the scale towards the overlapping
description. However, we emphasize that this extra information does
nothing regarding the decision between both collapsed models; it only
supports the acceptance of the third layered variant.

It is important to consider the above comparison together with the
results of Ref.~\cite{peixoto_model_2015}, which showed that the
overlapping variants of the SBM are seldom the best fit for the majority
of empirical networks used for that work, which contained no layer
information. As the example above shows, this assessment may change (at
least in principle) if any division among the edges can be assumed
\emph{a priori}. Therefore, for a fair assessment of the best generative
process, it is imperative to leverage all available information, in
particular the division into layers, or the existence of edge
covariates.

\section{Edges with real-valued correlates}\label{sec:real}

\begin{figure*}
  \centering
  \begin{tabular}{ccccc}
    \multicolumn{3}{c}{\includegraphics[width=.6\textwidth]{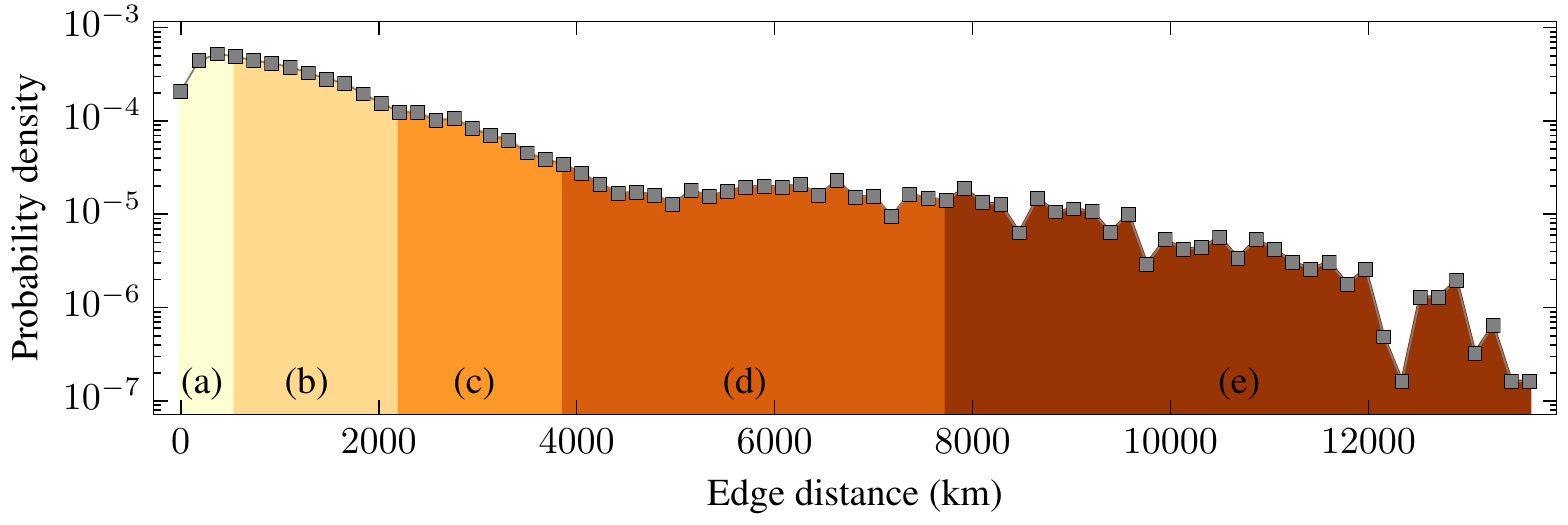}} &
    \multicolumn{2}{c}{\includegraphics[width=.4\textwidth]{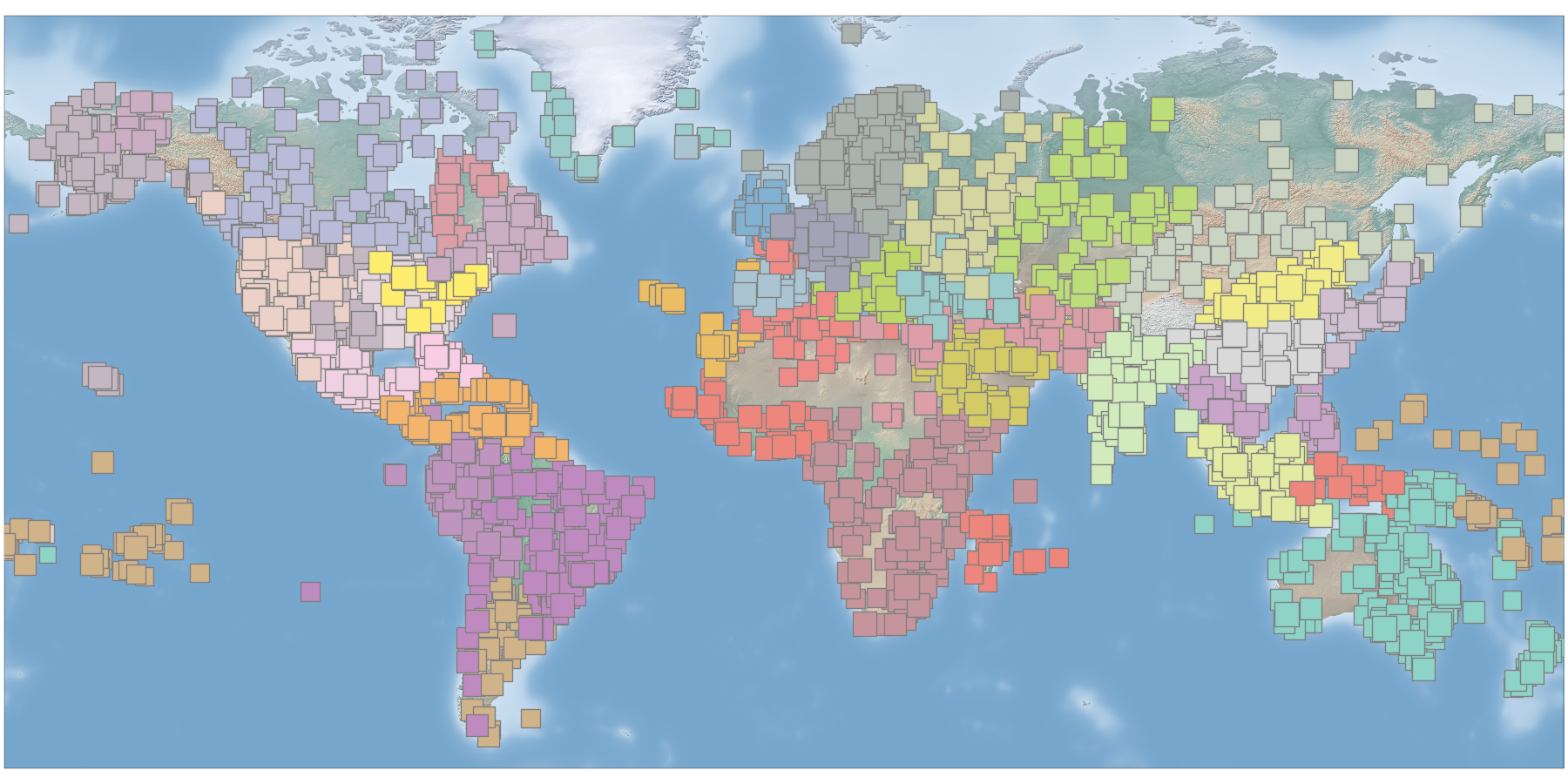}}\\
    \begin{overpic}[width=.2\textwidth]{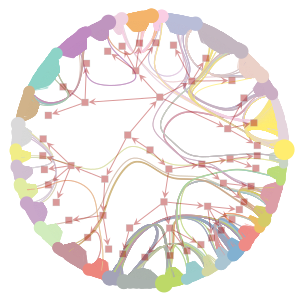}\put(0,0){\smaller (a)}\end{overpic} &
    \begin{overpic}[width=.2\textwidth]{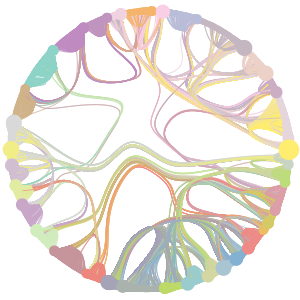}\put(0,0){\smaller (b)}\end{overpic} &
    \begin{overpic}[width=.2\textwidth]{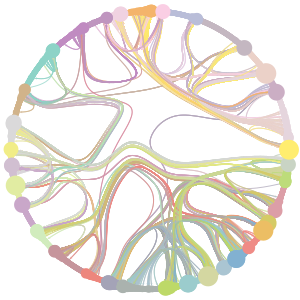}\put(0,0){\smaller (c)}\end{overpic} &
    \begin{overpic}[width=.2\textwidth]{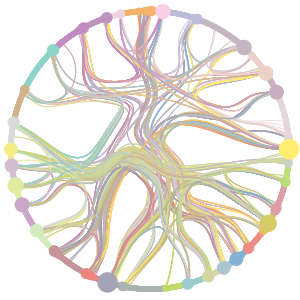}\put(0,0){\smaller (d)}\end{overpic} &
    \begin{overpic}[width=.2\textwidth]{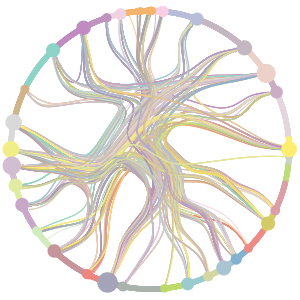}\put(0,0){\smaller (e)}\end{overpic}
  \end{tabular} \caption{(Color online) Global airport network of
  \url{openflights.org}. \emph{Top left:} Distribution of edge
  distances. The bins labeled from (a) to (e) correspond to the best
  division of the edges into layers according to the method described in
  the text.  \emph{Top right:} Spatial distribution of airports. The
  colors correspond to the division of the network into groups,
  according to the best fit of the DCSBM model with independent layers
  (the same color coding is used in the remaining
  panels). \emph{Bottom:} Individual layers of the DCSBM fit,
  corresponding to the bins in the top panel. The layout is the same as
  in Fig.~\ref{fig:physicians}. \label{fig:airport}}
\end{figure*}
\begin{figure*}
  \centering
  \begin{tabular}{ccccc}
    \multicolumn{5}{c}{\includegraphics[width=1\textwidth]{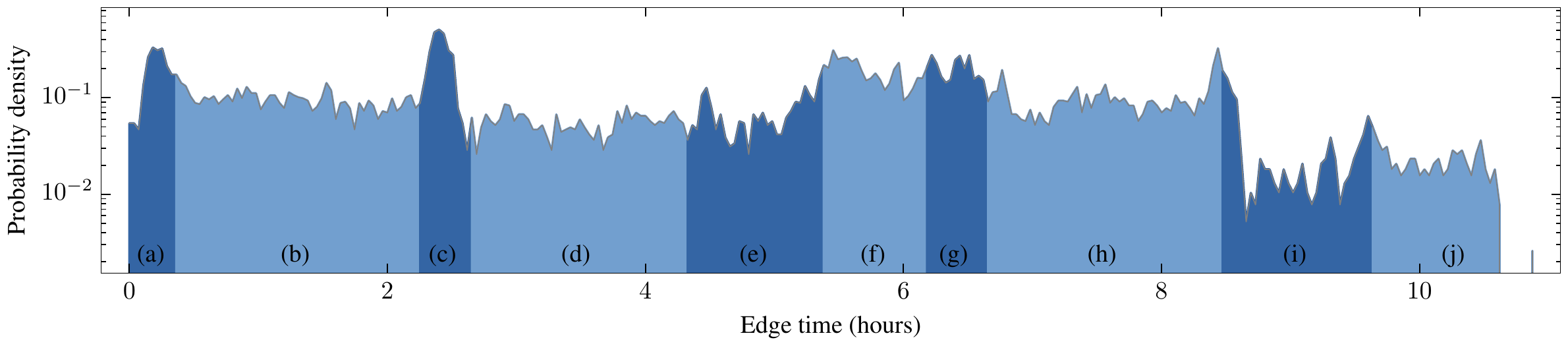}}\\
    \begin{overpic}[width=.215\textwidth]{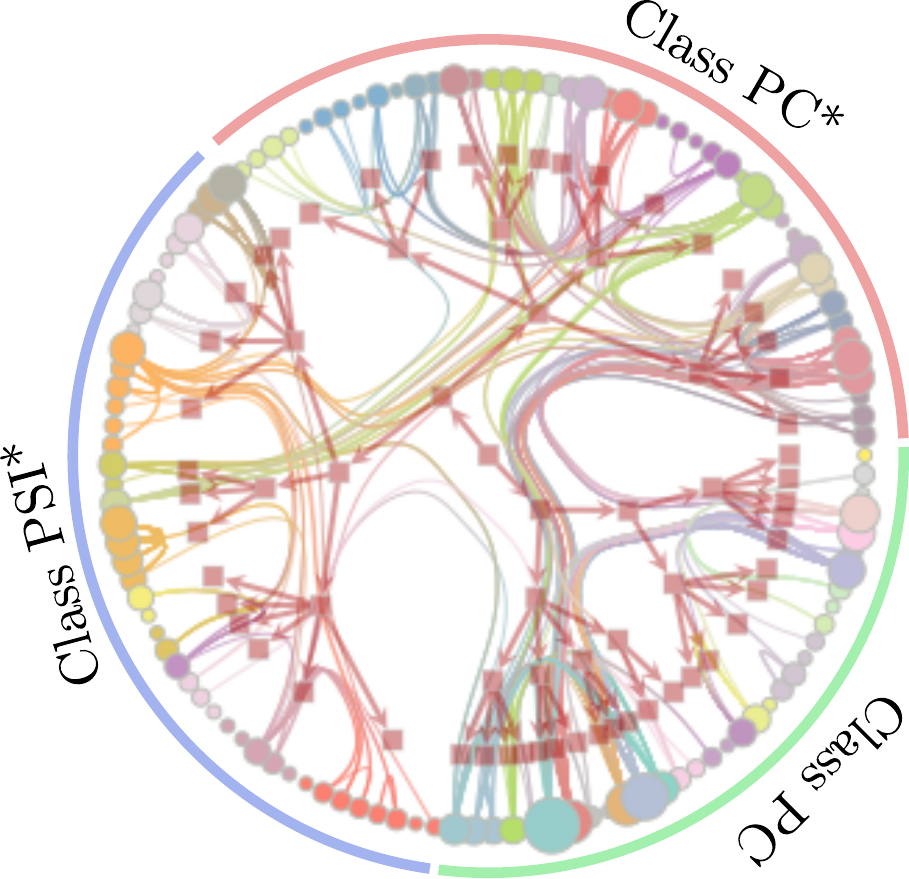}\put(0,0){\smaller (a)}\end{overpic} &
    \begin{overpic}[width=.2\textwidth]{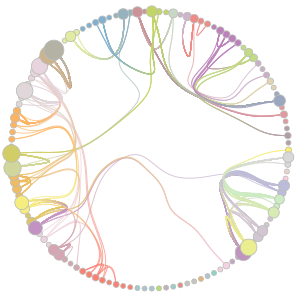}\put(0,0){\smaller (b)}\end{overpic} &
    \begin{overpic}[width=.2\textwidth]{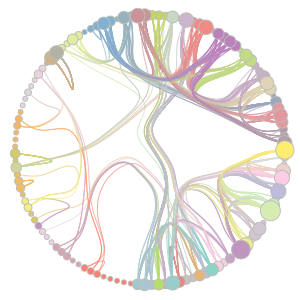}\put(0,0){\smaller (c)}\end{overpic} &
    \begin{overpic}[width=.2\textwidth]{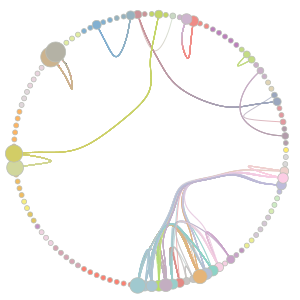}\put(0,0){\smaller (d)}\end{overpic} &
    \begin{overpic}[width=.2\textwidth]{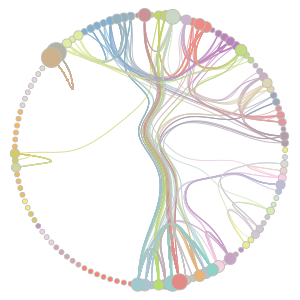}\put(0,0){\smaller (e)}\end{overpic} \\
    \begin{overpic}[width=.2\textwidth]{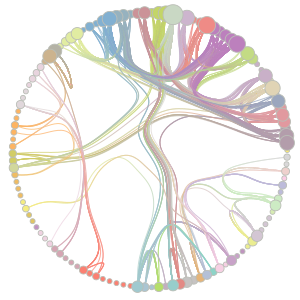}\put(0,0){\smaller (f)}\end{overpic} &
    \begin{overpic}[width=.2\textwidth]{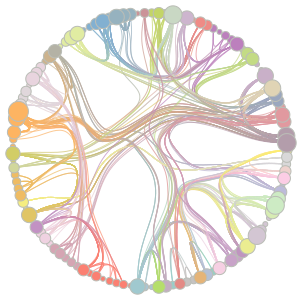}\put(0,0){\smaller (g)}\end{overpic} &
    \begin{overpic}[width=.2\textwidth]{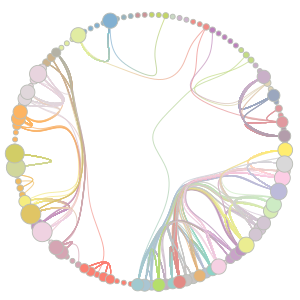}\put(0,0){\smaller (h)}\end{overpic} &
    \begin{overpic}[width=.2\textwidth]{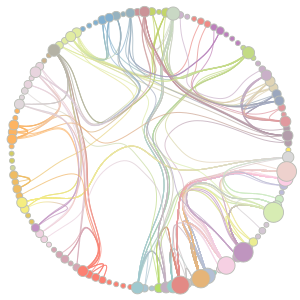}\put(0,0){\smaller (i)}\end{overpic} &
    \begin{overpic}[width=.2\textwidth]{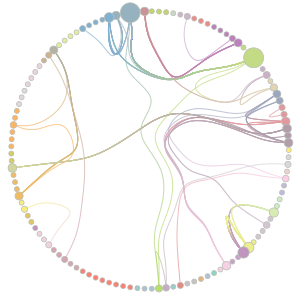}\put(0,0){\smaller (j)}\end{overpic}
  \end{tabular} \caption{(Color online) Proximity network between high-school
  students~\cite{fournet_contact_2014}\footnote{Retrieved from
  \url{http://sociopatterns.org}.}. \emph{Top:} Network activity
  (i.e. probability density of an edge being present) as a function of
  time, over a period of one day. The bins labeled from (a) to (j)
  correspond to the best division of the edges into layers according to
  the method described in the text. \emph{Bottom:} Individual layers of
  the DCSBM fit, corresponding to the bins in the top panel. The layout
  is the same as in Fig.~\ref{fig:physicians}.\label{fig:highschool} }
\end{figure*}

The models discussed so far are capable of generating data with discrete
values associated with the existing edges. However, in many important
situations the values associated with edges are real values,
corresponding to weights, distances, capacities, etc. Here we show how
the previous models can be straightforwardly adapted to these cases as
well, using a discretization approach. As before, we simply assume that
the graph is divided into $C$ discrete layers, however we ascribe to
each layer $l$ a real value $x_l$, randomly sampled from a PDF
$\rho(x)$, such that all edges in the same layer possess the same edge
correlate. In the case that all edges have a different correlate, we
will have $C=E$ layers. Like in Sec.~\ref{sec:info} we assume that the
layers themselves are grouped into bins $\{\ell\}$, with
$\{\theta\}_{\{\ell\}}$ being a shorthand for the possible set of
parameters of a layered SBM (with independent layers or edge covariates)
$\{G_\ell\}$ where each bin $\ell$ corresponds to an individual
layer. The whole PDF of the data generated in this manner becomes
\begin{multline}
  P(\{G_x\}|\{\theta\}_{\{\ell\}},\{\ell\}) =
  P(\{G_l\}|\{\theta\}_{\{\ell\}},\{\ell\}) \times \prod_l\rho(x_l),
\end{multline}
where the first term is given by Eq.~\ref{eq:p_ell}. The advantage of
this approach is that the overall correlate PDF $\rho(x_l)$ amounts to
constant multiplicative factor in the likelihood, independent of our
choice of bins, and therefore cannot influence either the maximum
likelihood estimate or the maximum of the posterior distribution, and
therefore for these purposes we can avoid specifying it altogether. This
contrasts with another generalization of the SBM for real-valued
covariates proposed in Ref.~\cite{aicher_learning_2014}, which requires
the exact form of the correlate distribution to be specified prior to
inference (on the other hand, the approach presented here is based the
discretization of the correlates into bins, whereas in
Ref.~\cite{aicher_learning_2014} no binning is necessary).

In order to choose the best number of layers, we maximize the posterior
$P(\theta_{\{\ell\}},\{\ell\}|\{G_x\})$, which involves the priors of
the SBM parameters, as well as for the bins $\{\ell\}$, as given by
Eq.~\ref{eq:lbins}. Therefore, both the number and the boundary
positions of the bins can be determined in a nonparametric manner, based
only on the data.

As an example we consider the global airport network as collected by
\url{openflights.org}. This is a directed multigraph, where the $N=3253$
nodes are airports and the $E=67154$ edges represent existing
flights. Since the position of the airports is known, we can
characterize the edges by their geodesic distance, which we treat as a
covariate. In applying the DCSBM with independent layers, using the
method outlined above to find the optimal binning of the distances, we
find a division into $B=34$ groups, and $M=5$ distance bins, as shown in
Fig.~\ref{fig:airport}. When inspecting the spatial distribution of
airports, we observe that the obtained groups correspond to fairly
contiguous geographical regions (see Fig.~\ref{fig:airport}, top
right). The distribution of edges across the layers reveal a
hierarchical organization strongly correlated with flight distance: The
first layer captures local ``intra-groups'' with relatively short
distance, whereas the upper layers capture increasingly ``inter-groups''
flights with longer distances. The nodes with large degree tend to be
those that belong to multiple layers,
i.e. major airport hubs that service both short and long-distance
flights.

\section{Time-varying networks}\label{sec:time}

Temporal networks can be viewed as a special case of networks with
real-valued edge correlates representing their existence at a specific
time, $x_i=t_i$, and hence we can use the same approach as in the
previous section\footnote{Other formulations of temporal networks are
possible. For instance, one could attribute to each edge a tuple
$\vec{x}_i=(t^b_i, t^e_i)$, containing a creation and deletion time,
respectively. The approach presented here can be adapted to such a
multivariate case in a straightforward manner, by using multidimensional
bins.}. By using the different model versions presented in this work,
different types of temporal patterns can be captured. In all cases, by
separating the network into time-bins, it is assumed that inside each
bin the edges are placed between the groups in a random fashion,
conditioned only on the group membership of the receiving nodes. When
using the SBM with edge covariates, the nodes are assumed to belong to
all time layers, and as such can receive edges at all times, depending
only on the activity of the entire group at any give time. On the other
hand, the version with independent layers allows for a individualized
placement of the nodes into the layers (independently of their group
membership) such that their activity may be separately regulated. The
activity inside each layer can be even more fine-tuned in the
degree-corrected model with independent layers, since the degree of each
node at each time window is separately specified. In all these examples,
the group memberships are forced to be stable in time. This can be
changed by using an overlapping SBM~\cite{peixoto_model_2015}, where the
group memberships (which are in this case attributes of the
\emph{half-edges} of the graphs) can change arbitrarily in time. As
before, given some empirical observation, the most appropriate model
choice is the one with the minimum description length.

The discretization approach presented here is similar in spirit to the
detection of ``change points'' in
networks~\cite{peel_detecting_2014}. Since it is assumed that inside
each time window the edges are placed in a manner that is independent of
their time relative to one another, the most appropriate time binning is
the one that partitions the time series in such a way that inside each
time window the large-scale network structure does not change
significantly. The interface between two bins can therefore be
interpreted as change points where the large-scale structure has changed
in a measurable and statistically significant way.

Here we show an application of this method to a time-resolved proximity
network between $N=126$ high-school students, recorded over a period of
four days in 2011~\cite{fournet_contact_2014}, of which we isolated only
the first day to simplify the analysis. In this experiment, volunteering
students wore proximity sensors during school hours, which recorded an
edge and its time if two students were below a distance threshold for a
pre-specified amount of time. If we apply the DCSBM with independent
layers to this dataset (again providing a better fit), the best
partition is found for $B=33$ groups, and the whole time series was
divided into $M=10$ periods, as can be seen in
Fig.~\ref{fig:highschool}. The hierarchical partition is in accordance
with the existence of three classes, as can be seen in the first levels
of the hierarchy. Each period marks a region in time where a distinct
large-scale structure is observed. These periods alternate between those
with high activities and those with a relative quiescence, presumably
representing breaks (with many edges between classes, and a perceived
synchrony between the PC and PC$^*$ classes) and class periods (with few
edges between classes), respectively, although this information is not
available in the dataset.

\begin{figure}
  \centering \includegraphics[width=\columnwidth]{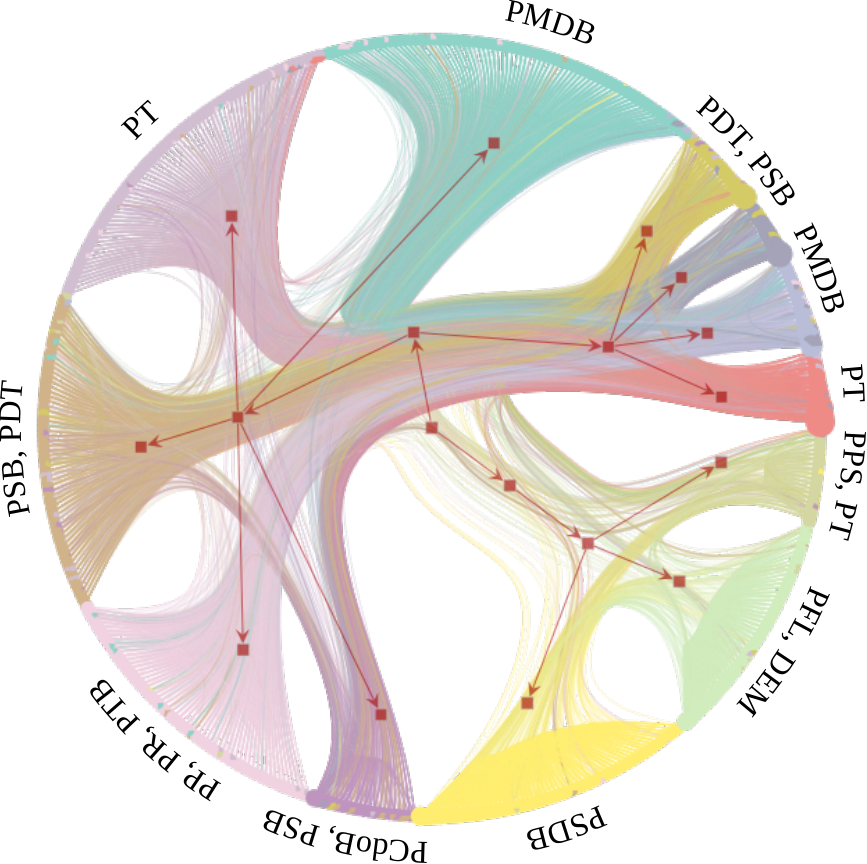}
  \includegraphics[width=\columnwidth]{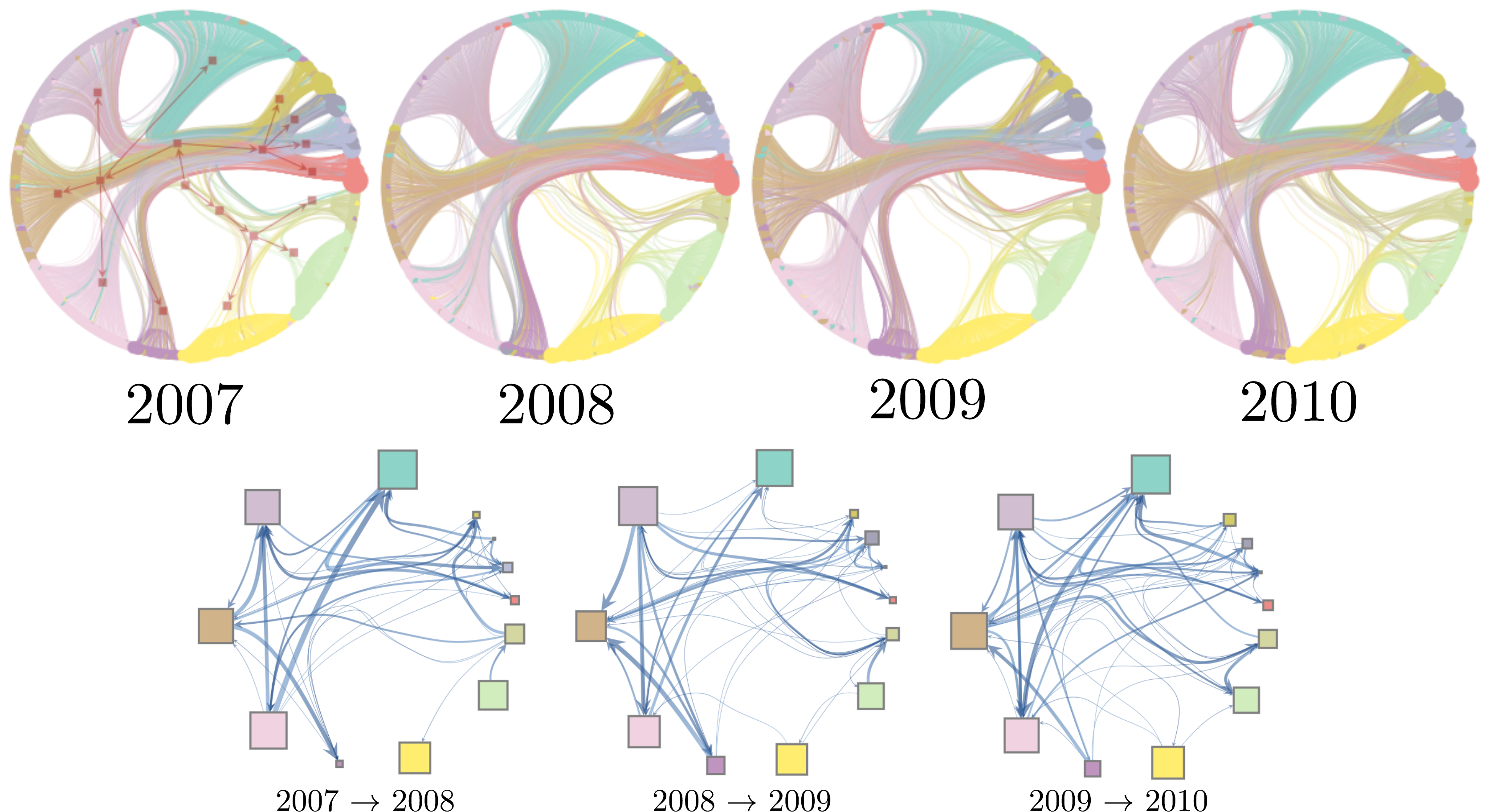} \caption{(Color online) Network
  of vote correlations among federal deputies of the Brazilian national
  congress in the four-year term from 2007 to 2010. The top panel shows
  the $B=12$ division obtained by fitting an overlapping DCSBM with
  independent layers, with all layers collapsed into one figure. The
  group labels correspond to the predominant parties inside each
  group. The individual layers can be seen in the middle panel. The
  bottom panel shows the flows of deputies between each group after each
  year. The edge thickness corresponds to the amount of deputies, with
  the largest flow corresponding to 10 deputies, and the smallest 1
  deputy. \label{fig:camara_flow}}
\end{figure}

In the above example, the best fit was obtained for a nonoverlapping
SBM, implying that the group memberships remain stable in time. However,
in some situations, movements between groups can be inferred. As an
example, we return to the network of vote correlations of the Brazilian
national congress. Differently from before, now we inspect a single
four-year term from 2007 to 2010, and we separate each year into one
layer, yielding a network with $N=475$ nodes and $E=9053$ edges in
total. In this case, a best fit is obtained for an \emph{overlapping}
DCSBM with independent layers and $B=12$ groups, as seen in
Fig.~\ref{fig:camara_flow}. The hierarchical division clearly separates
between a center-left government coalition (the largest topmost branch)
and the right-wing opposition (the smallest topmost branch). In the
government branch, we observe the existence of many ``peripheral''
deputies, which are not strongly correlated with each other, and instead
are aligned with smaller groups of more connected nodes, which are
divided mostly along party lines. This property is weakened in the later
years of the term, as more edges are observed between peripheral
deputies.  The overlapping structure found is correlated strongly with
the layered divisions, such that by observing only one layer in
isolation, no overlaps are present. Therefore, a fraction of the
deputies seem to completely change their alignment patterns in
successive years, as shown in the bottom of
Fig.~\ref{fig:camara_flow}. The flow between groups is mostly confined
to either the government or opposition groups, with the majority of the
activity occurring inside the government faction. Although some deputies
did change their party affiliation during this period, the observed
flows seem mostly uncorrelated with this, and instead appears to show a
more fine-grained alignment between deputies that is not uniquely
defined by their party membership.

\section{Conclusion}\label{sec:conclusion}

We presented a framework for the nonparametric inference of mesoscale
structures in layered, edge-valued and time-varying networks, based on a
variety of modifications of the stochastic block model, incorporating
features such as hierarchical structure, degree-correction, and
overlapping groups. These models were formulated in a Bayesian setting,
that allows the identification of the most appropriate model variant
based on statistical evidence, corresponding to a principled balance
between model complexity and quality of fit.

We have identified an important pitfall when analyzing network data with
layered structure, where the inclusion of many layers that are
uncorrelated with the mesoscale structure can obstruct its
identification. This problem cannot be neglected if the number of layers
becomes large, as in the case of temporal or edge-value networks where
the layers correspond to arbitrary bins of the edge covariates. We
expect this problem to affect also non-statistical methods based on
modified modularity maximization~\cite{mucha_community_2010,
de_domenico_mathematical_2013,bazzi_community_2014,sarzynska_null_2014,macmahon_community_2015},
as well as flow compression~\cite{de_domenico_identifying_2015} and
non-negative tensor factorization~\cite{gauvin_detecting_2014}. In our
setting, we have shown how this can be completely avoided by comparing
the inferred model with a null model that assumes that the layers are
uncorrelated, or with a coarse-grained version that condenses
uncorrelated layers into bins.

We also showed how this framework can be extended in a straightforward
manner to networks with real-valued attributes on the edges, and
temporal networks. The proposed methodology is capable of identifying
specific scales --- both of the edge values and in time --- where the
mesoscale structure does not change significantly, enabling the
identification of the most appropriate coarse-graining of the network in
discrete layers, as well as the detection of ``change points'' of the
network structure.

The unsupervised inference of the most parsimonious layered model, as
well as the appropriate granularity of the layers, based solely on
statistical evidence and requiring no \emph{ad hoc} parameters, provides
a principled and robust method to analyze multilayer, temporal and
edge-valued network data. This approach is likely to be directly useful
in a variety of tasks, such as the nonparametric modeling of
correlation networks~\cite{bazzi_community_2014}, the prediction of
missing valued edges~\cite{guimera_network_2013,
rovira-asenjo_predicting_2013}, the identification of relevant time
scales in temporal networks~\cite{masuda_temporal_2013}, and its
relation to dynamical processes taking place on
them~\cite{gemmetto_mitigation_2014, voirin_combining_2015}, among many
others.

\begin{acknowledgments}
I would like to thank Ricardo Marino for compiling and kindly providing
the voting data of the chamber of deputies in the Brazilian national
congress, Roger Guimerà and Marta Sales-Pardo for insightful
discussions, as well as Mason Porter and Martin Rosvall for helpful
comments on the manuscript. This work was funded by the University of
Bremen, under the program ZF04.
\end{acknowledgments}
\appendix

\section{Multigraphs}\label{app:multigraphs}

For multigraphs, we need to consider that parallel edges that belong to
the same layer are indistinguishable. Hence the likelihoods of
Eq.~\ref{eq:l_covariates} must be corrected
to read
\begin{multline}
  P(\{G_l\}|\{\theta\}) =
  P(G_c|\{\theta\})\prod_{r\leq s}\frac{\prod_lm_{rs}^l!}{m_{rs}!}  \times  \\
  \frac{\prod_{i>j}A_{ij}!}{\prod_{i>j,l}A^l_{ij}!}
  \frac{\prod_iA_{ii}/2!}{\prod_{i,l}A^l_{ii}/2!}.
\end{multline}
The last term does not depend on the SBM parameters. Therefore, when
doing inference, the difference amounts to multiplicative constant which
does not alter the position of the most likely network partition, and
thus could in principle be discarded. However, this difference is
important when comparing models with a different number of layers, as
will be done below.

For the independent layers model, it suffices to use the appropriate
multigraph likelihood in each layer, as is given in
Refs.~\cite{peixoto_entropy_2012,peixoto_model_2015}.

Likewise, when considering the null model of Sec.~\ref{sec:info}, the
existence of parallel edges must also be accounted for. Therefore
Eq.~\ref{eq:null} must be modified to read
\begin{multline}
  P(\{G_l\}|\{\theta\},\{E_l\}) =
  P(G_c|\theta)\times \frac{\prod_lE_l!}{E!} \times  \\
 \frac{\prod_{i>j}A_{ij}!}{\prod_{i>j,l}A^l_{ij}!} \times
 \frac{\prod_iA_{ii}/2!}{\prod_{i,l}A^l_{ii}/2!}.
\end{multline}
In the case of binned layers, it must be analogously modified to read
\begin{multline}
  P(\{G_l\}|\theta_{\{\ell\}},\{\ell\})  =
  P(\{G_\ell\}|\theta_{\{\ell\}})\times \prod_\ell\frac{\prod_{l\in\ell}E_{l}!}{E_\ell!} \times \\ 
  \prod_{i>j}\prod_{\ell}\frac{A^{\ell}_{ij}!}{\prod_{l\in \ell}A_{ij}^l!} \times
  \prod_{i}\prod_{\ell}\frac{A^{\ell}_{ii}/2!}{\prod_{l\in \ell}A_{ii}^l/2!}.
\end{multline}

\section{Directed graphs}\label{app:directed}

Directed graphs represent straightforward modifications of the models
presented in the main text. For the collapsed likelihoods and priors, we
refer the Refs.~\cite{peixoto_entropy_2012,peixoto_model_2015}.

For the model with edge covariates and the possibility of multiple
edges, the total likelihood of Eq.~\ref{eq:l_covariates} becomes simply
\begin{multline}
  P(\{G_l\}|\{\theta\}) =
  P(G_c|\{\theta\})\prod_{rs}\frac{\prod_lm_{rs}^l!}{m_{rs}!} \times
  \frac{\prod_{ij}A_{ij}!}{\prod_{ij,l}A^l_{ij}!}.
\end{multline}
And again, for the independent layers model, it suffices to use the
appropriate directed likelihood in each layer, as is given in
Refs.~\cite{peixoto_entropy_2012,peixoto_model_2015}.

Likewise, when considering the null model of Sec.~\ref{sec:info}, for
directed graphs (with possible multiple edges) Eq.~\ref{eq:null} must be
modified to read
\begin{multline}
  P(\{G_l\}|\{\theta\},\{E_l\}) =
  P(G_c|\theta)\times \frac{\prod_lE_l!}{E!} \times
 \frac{\prod_{ij}A_{ij}!}{\prod_{ij,l}A^l_{ij}!},
\end{multline}
and in the case of binned layers,
\begin{multline}
  P(\{G_l\}|\theta_{\{\ell\}},\{\ell\})  =
  P(\{G_\ell\}|\theta_{\{\ell\}})\times \prod_\ell\frac{\prod_{l\in\ell}E_{l}!}{E_\ell!} \times  \\
  \prod_{ij}\prod_{\ell}\frac{A^{\ell}_{ij}!}{\prod_{l\in \ell}A_{ij}^l!}.
\end{multline}

\section{Model selection for overlapping groups}\label{app:overlap}

In the case the SBM with overlapping groups, we need to specify a
generative process for the overlapping partition into $B$ groups
$\{\vec{b}_i\}$, and the collapsed (or layer-specific) labeled
degree-sequence $\{\vec{k}_i\}$ (or $\{\vec{k}^l_i\}$).

To generate the overlapping partition into groups, we use the
hierarchical process described in detail in
Ref.~\cite{peixoto_model_2015}, already described in the main text
adapted to the generation of the layer-membership matrix $\{z_{il}\}$,
which yields
 $\mathcal{L}_p=-\ln P(\{\vec{b}_i\})$ given by
\begin{align}\label{eq:o_lp}
  \mathcal{L}_p = \ln{\textstyle\multiset{D}{N}} + \sum_d \ln {\textstyle\multiset{{B \choose d}}{n_d}} + \ln N! - \sum_{\vec{b}}\ln n_{\vec{b}}!,
\end{align}
where $D \le B$ is the maximum mixture size $d$. The case without group
overlaps amounts to $D=1$, reducing it to Eq.~\ref{eq:lp}.

The \emph{collapsed} overlapping degree sequence can be generated with a
similar Bayesian process, described also in
Ref.~\cite{peixoto_model_2015}, that yields a description length
$\mathcal{L}_{\kappa} = -\ln P(\{\vec{k}_i\})$ given by
\begin{equation}\label{eq:o_lkappa}
  \mathcal{L}_{\kappa} = \sum_r\ln{\textstyle\multiset{m_r}{e_r}} + \sum_{\vec{b}}\min\left(\mathcal{L}^{(1)}_{\vec{b}}, \mathcal{L}^{(2)}_{\vec{b}}\right).
\end{equation}
with
\begin{align}
  \mathcal{L}^{(1)}_{\vec{b}} &= \sum_r\ln{\multiset{n_{\vec{b}}}{e^r_{\vec{b}}}}, \\
  \mathcal{L}^{(2)}_{\vec{b}} &= \sum_{r\in\vec{b}}\ln\Xi_{\vec{b}}^r + \ln n_{\vec{b}}! - \sum_{\vec{k}} \ln n^{\vec{b}}_{\vec{k}}!.
\end{align}
where $\ln\Xi_{\vec{b}}^r \approx 2\sqrt{\zeta(2)e_{\vec{b}}^r}$. For the
case without overlaps this reduces to Eq.~\ref{eq:lkappa_D1}. The
edge-specific overlapping degree sequence is obtained according to
Eq.~\ref{eq:lkappa}.

\bibliography{bib}
\end{document}